# How to Distinguish AI-Generated Images from Authentic Photographs

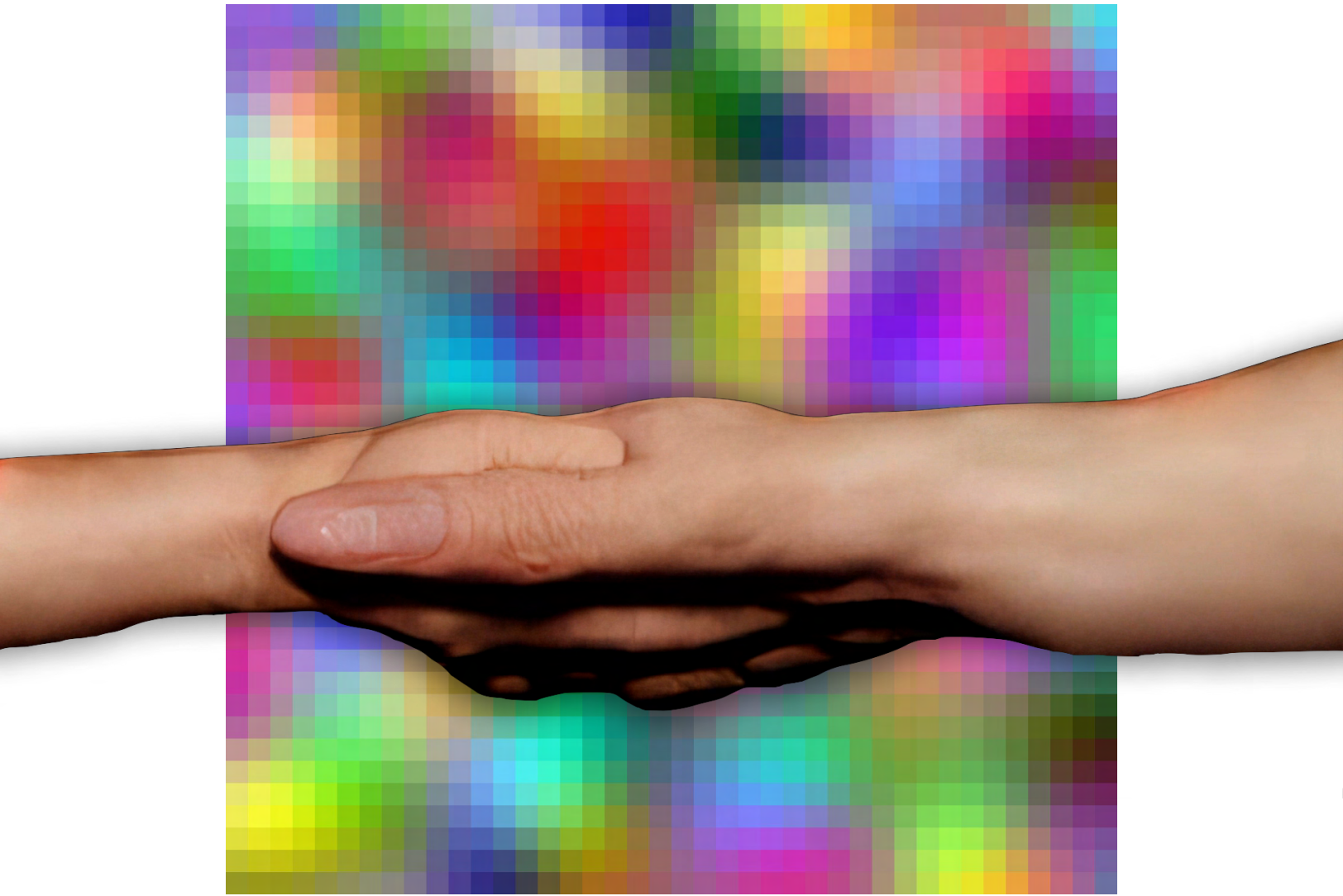

Negar Kamali, Karyn Nakamura, Angelos Chatzimparmpas,
Jessica Hullman, and Matthew Groh

Northwestern University
June 2024

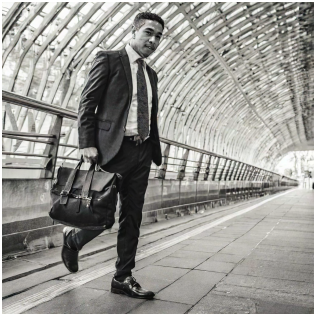 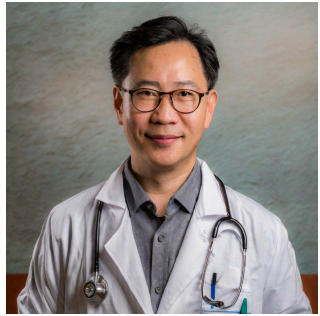 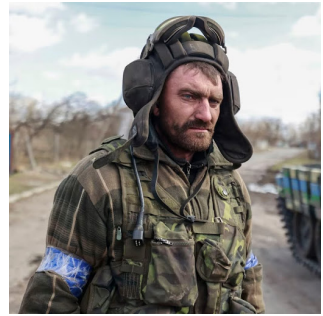

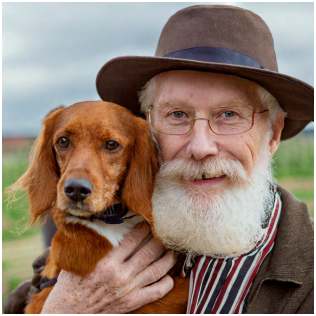 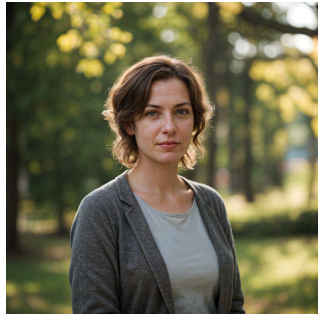 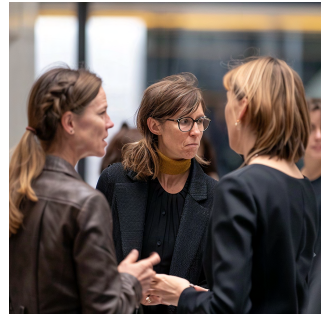

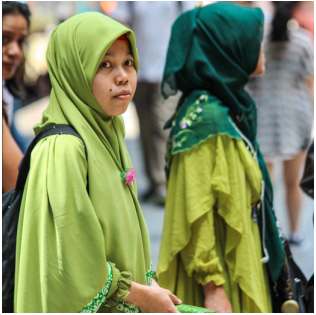 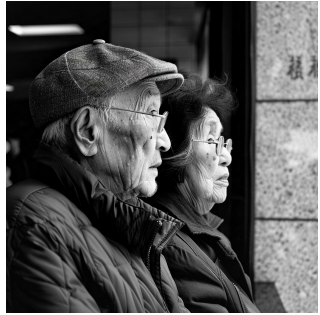 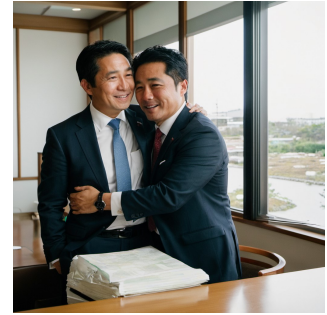

Two of these images are real and the rest have been generated by AI. Can you spot the real ones? Look closely!

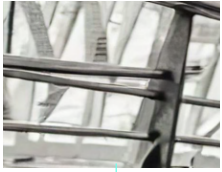

extra nonfunctional pole

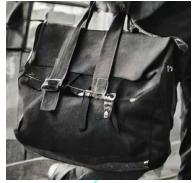

asymmetrical buckles

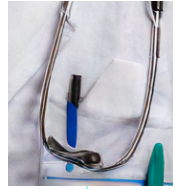

merged stethoscope earpieces

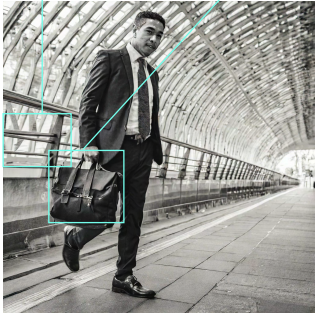

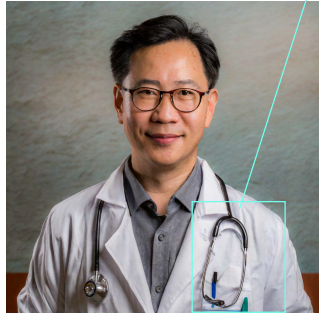

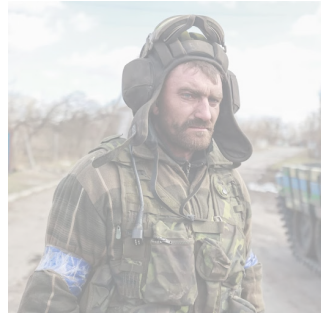

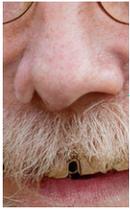

visual artifact between teeth

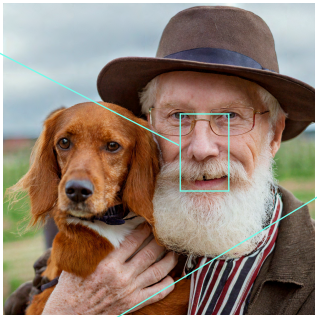

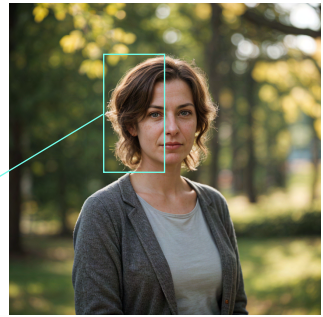

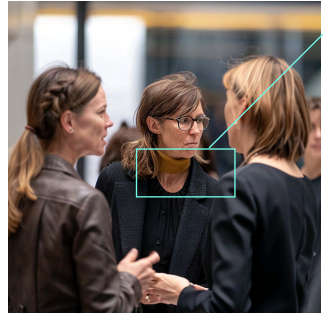

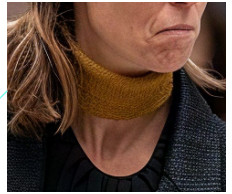

atypical detached collar design

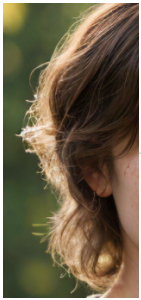

extra wispy hair

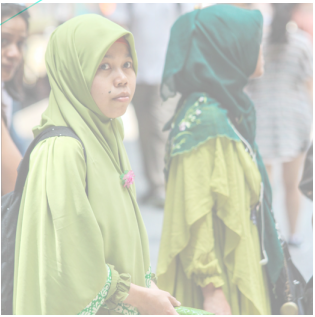

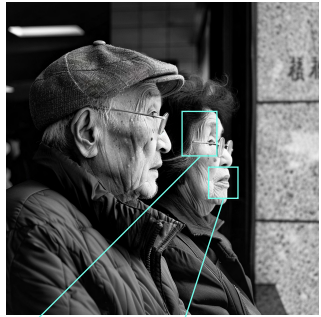

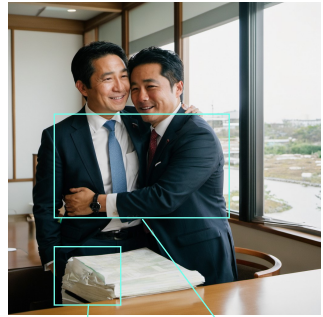

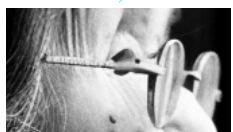

missing eye detail

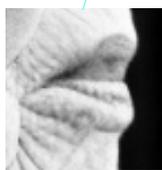

overlapping of teeth and mouth

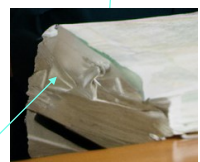

blurred resolution artifact

hugging is uncommon in Japanese (and most Asian) cultures, especially in a professional setting

## Abstract

The high level of photorealism in state-of-the-art diffusion models like Midjourney, Stable Diffusion, and Firefly makes it difficult for untrained humans to distinguish between real photographs and AI-generated images. In order to address this problem, we designed a guide to help readers develop a more critical eye towards identifying artifacts, inconsistencies, and implausibilities that often appear in AI-generated images. The guide is organized into five categories of artifacts and implausibilities: anatomical, stylistic, functional, violations of physics, and sociocultural. For this guide, we generated 138 images with diffusion models, curated 9 images from social media, and curated 42 real photographs. These images showcase the kinds of cues that prompt suspicion towards the possibility an image is AI-generated and why it's often difficult to draw conclusions about an image's provenance without any context beyond the pixels in an image. Human-perceptible artifacts are not always present in AI-generated images, but this guide reveals artifacts and implausibilities that often emerge. By drawing attention to these kinds of artifacts and implausibilities, we aim to better equip people to distinguish AI-generated images from real photographs in the future.

## Acknowledgements

We thank Robert Pozen, the Laboratory for Analytic Sciences at North Carolina State University, and Kellogg School of Management for generous support.

# Introduction

# 01. Anatomical Implausibilities                                      15

# 02. Stylistic Artifacts                                              23

# 03. Functional Implausibilities                                      30

# 04. Violations of Physics                                            38

# 05. Sociocultural Implausibilities                                   43

# Appendix                                                             51

# Introduction

The real photos on page 2 are a photo of a Ukrainian serviceman standing in a street in Lukyanivka, Kyiv region, Ukraine, on March 27, 2022 from The Atlantic and a photo taken at a bus stop in Hong Kong on a Canon EOS Rebel by Oleg S found on Flickr. If you'd like to test your skills more, visit our website: https://detectfakes.kellogg.northwestern.edu.

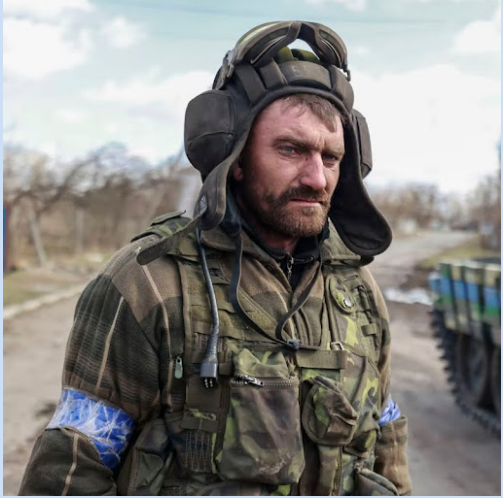

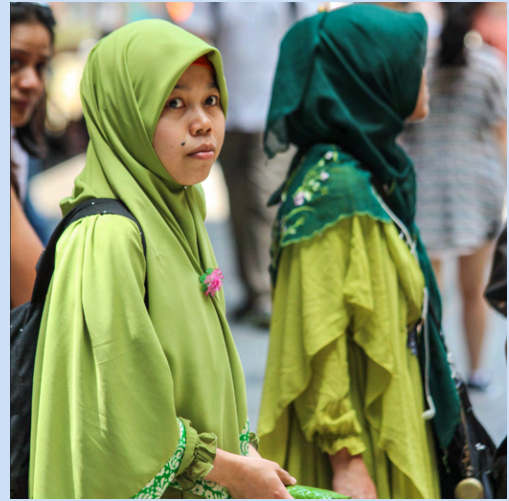

*Faces of Ukraine,* April 2022 (The Atlantic)
\* The blue tape armband that may look suspicious, is a form of identification adopted by the Ukrainian forces to distinguish them from Russian troops.

Matching colors by Oleg S, November 2016 (Flickr)

All other photos on page 2 were generated using commercial (Midjourney and Firefly) and open-source (Stable Diffusion) AI image generation tools. In 2024, it is becoming difficult to distinguish whether some images are real or generated by AI. On first glance, many AI-generated images appear real. However, a second look often reveals a number of artifacts and implausibilities that can help a critical viewer identify that an image was generated by AI. Given the ease with which people can generate photorealistic, provocative images to incite reactions and mislead, it is important to ask oneself, could this image have been generated by AI?

We organized this 2024 guide across 5 high level categories in which artifacts and implausibilities emerge in AI-generated images: **Anatomical Implausibilities, Stylistic Artifacts, Functional Implausibilities, Violations of Physics,** and **Sociocultural Implausibilities**. However it is not always possible to readily identify artifacts and implausibilities in images, especially portrait images. Likewise, authentic photographs sometimes contain elements that appear implausible or like a visual artifact. The goal of this guide is to help you develop a sharper eye for visual inconsistencies and calibrate your intuition for whether an image is AI-generated, authentic, or too ambiguous to know without further information.

## 01. Anatomical Implausibilities

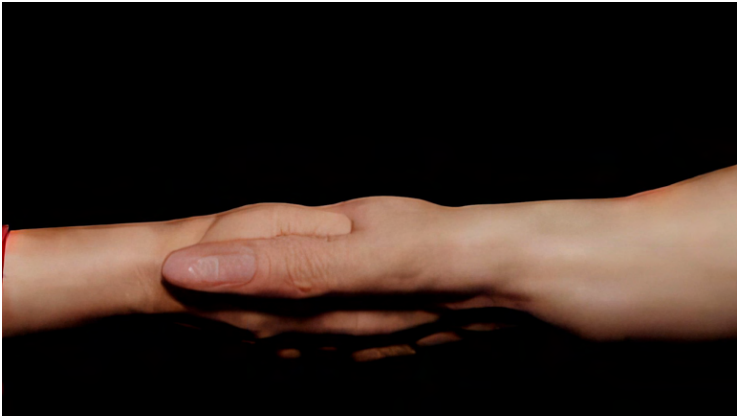

## 02. Stylistic Artifacts

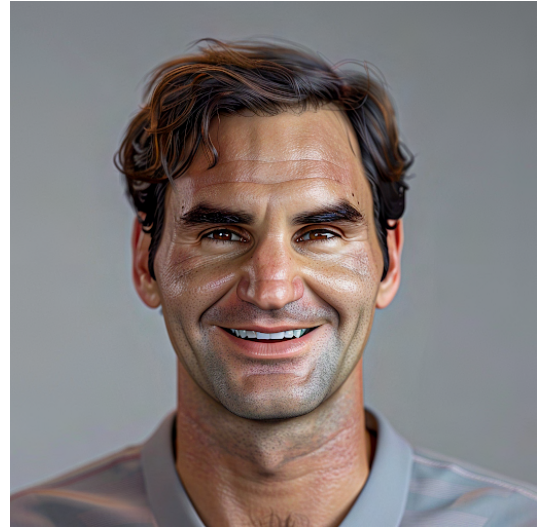

## 03. Functional Implausibilities

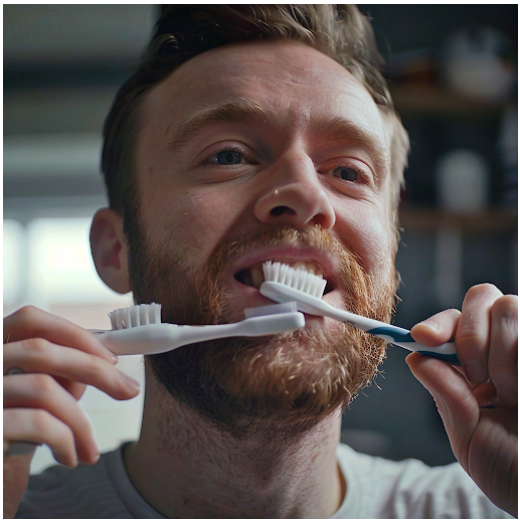

## 04. Violations of Physics

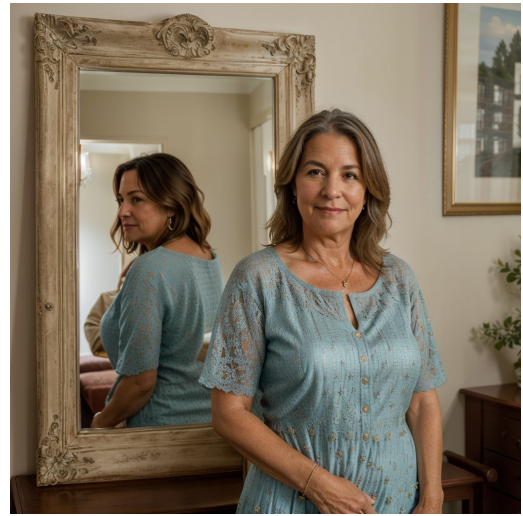

## 05. Sociocultural Implausibilities

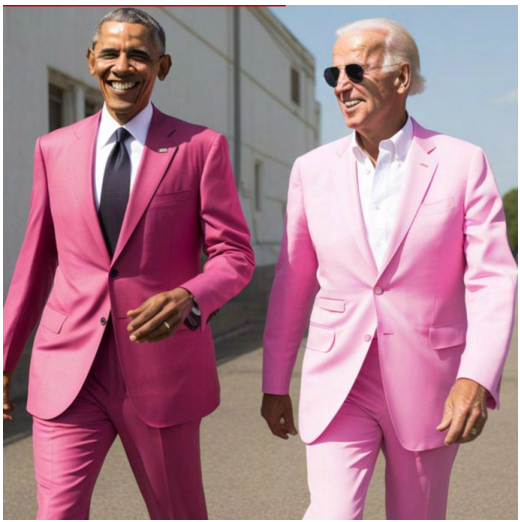

## 0.2 Brief Primer on AI Image Generation

This how-to guide focuses on images that have been entirely generated by AI via diffusion models. Photorealistic AI-generated images began to emerge with Generative Adversarial Networks (GANs) in 2014. The extreme photorealism of GAN portraits entered popular culture after websites like This Person Does not Exist went viral. The images below on the left reveal the progress towards high-resolution photorealism of face portraits by GANs from 2014 to 2021.

In 2024, diffusion models are the state-of-the-art for generating highly expressive and controllable photorealistic images. The images below on the right show images with faces generated by GANs (generated.photos, StyleGAN, EG3D) and diffusion models (DALL-E, Midjourney, and Stable Diffusion).

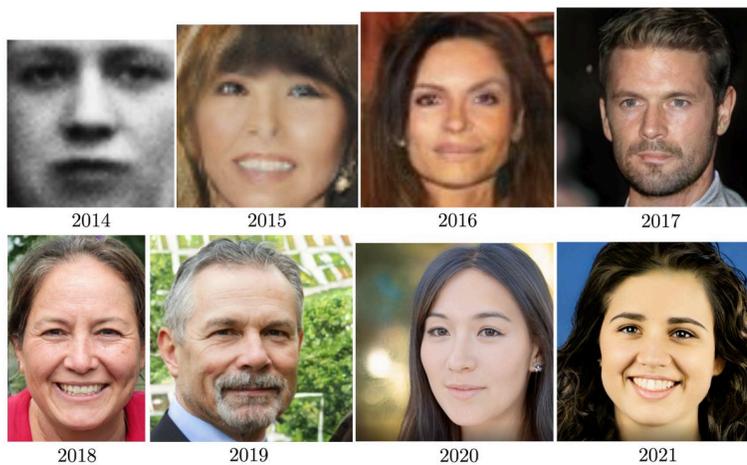

7.5 years of GAN progress on face generation by Tamay Besiroglu

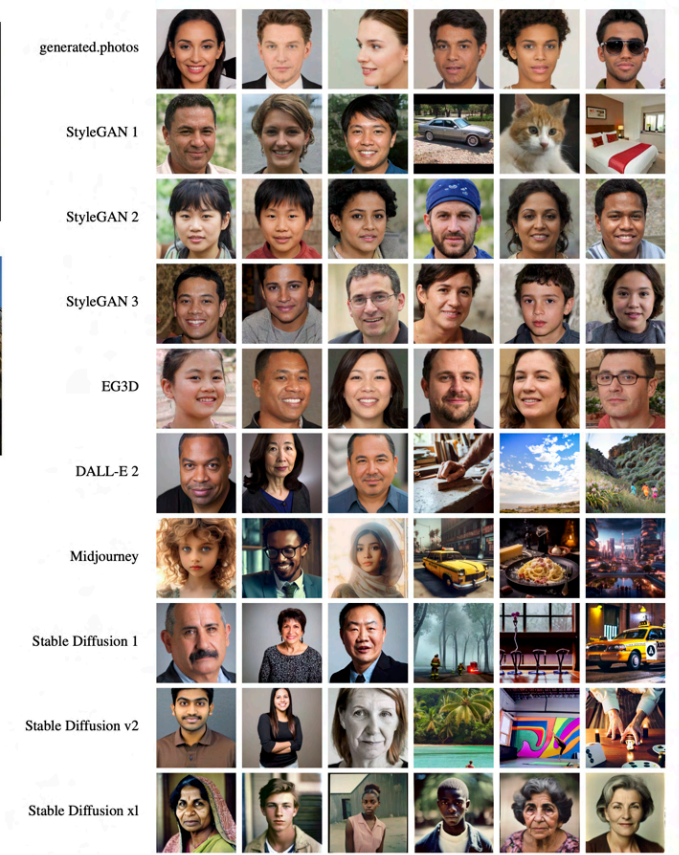

Finding AI-Generated Faces in the Wild, Porcile, G. J. A., Gindi, J, Mundra, S, Versus, J. R., & Farid, H. (2024)

**Introduction**

Diffusion models are trained by adding structured noise (called Gaussian noising) to an image as shown in the diagram below and learning to remove noise iteratively to return to an image. Once a model is trained, it generates an images by starting with an image of random noise and refining it into a coherent picture through gradually removing noise.

Diffusion model platforms like Midjourney, Stable Diffusion, Firefly, and DALL-E allow anyone to create AI images through a prompt interface that calls this process.

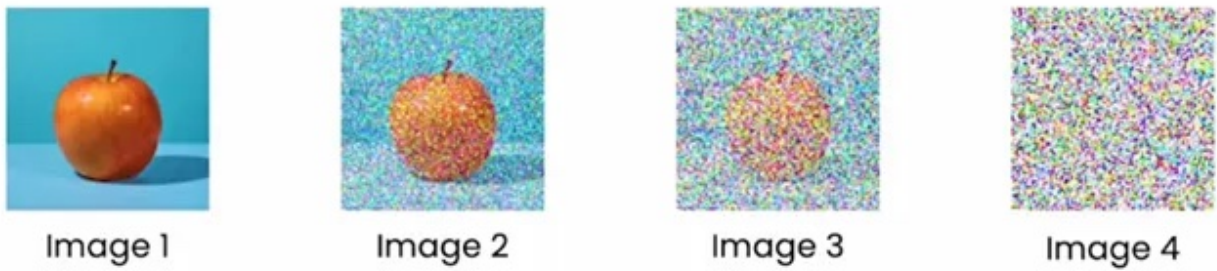

Diagram of the training process of a diffusion model. Noise is added to an image, which will then be used to train the model to take a noisy image (A) as input and then produce a slightly less noisy image (B) as output. From Andrew Ng's Generative AI for Everybody Course (Coursera)

Throughout this guide we have generated images with Midjourney (V5 and V6), Stable Diffusion (1.5, SDXL, and 3), and Firefly (version 2) to demonstrate the capabilities of diffusion models and reveal the artifacts and implausibilities that often emerge. While there are other AI image generation platforms, we use Midjourney and Firefly for their ease of use and large user community (especially around Midjourney) as well as Stable Diffusion that allows for more flexibility, controlled generation, and generation of content that would be moderated under other models such as deepfakes of Joe Biden and Donald Trump. We have also curated AI-generated images that have gone viral online and real photos as part of the guide, which will be indicated with a link to their source.

On the next page, we present a gallery of AI-generated images we have created, categorized by 4 types of images: portraits, full-body, posed group, and candid group.

## Portraits

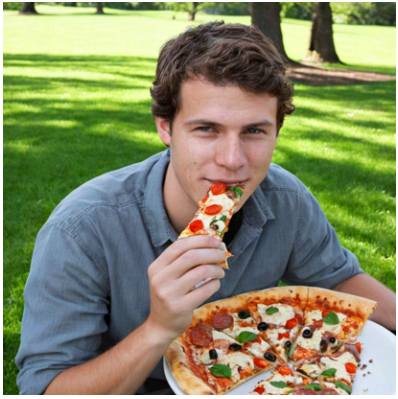
Stable Diffusion

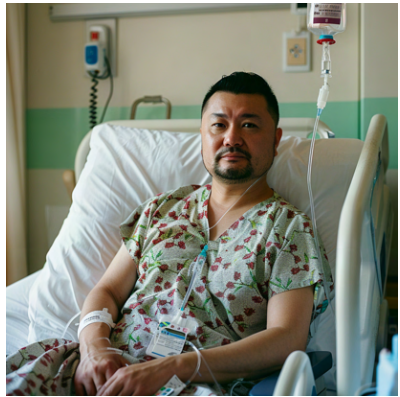
Midjourney

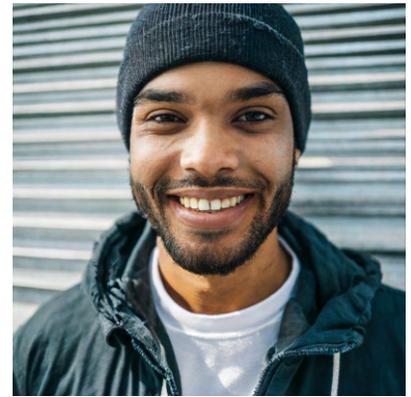
Firefly

## Full-body

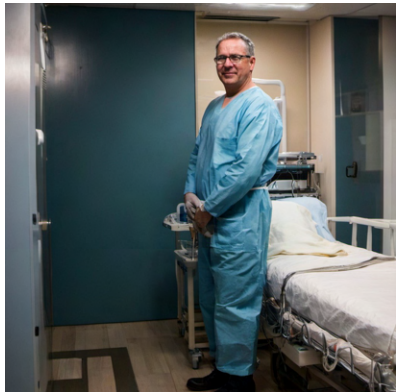
Stable Diffusion

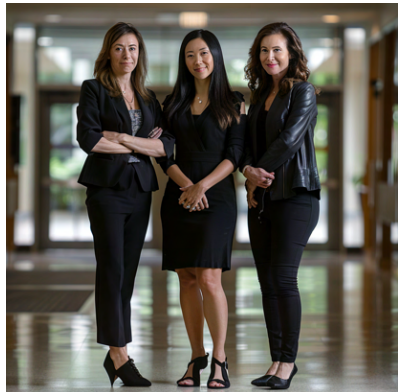
Midjourney

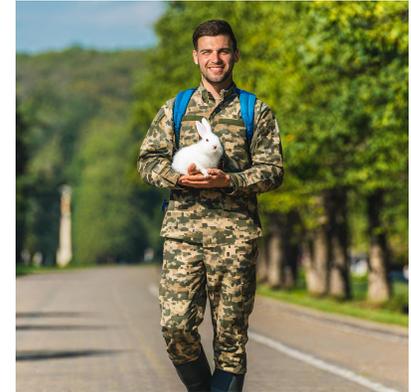
Firefly

## Posed Group

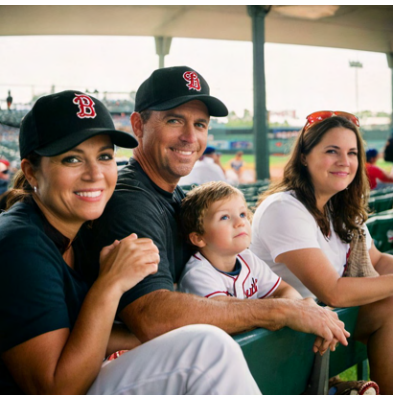
Stable Diffusion

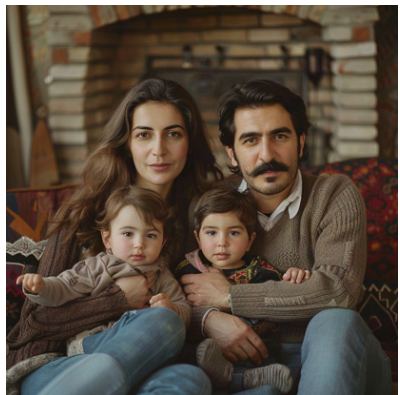
Midjourney

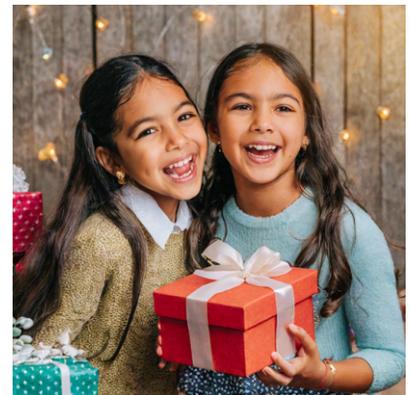
Firefly

## Candid Group

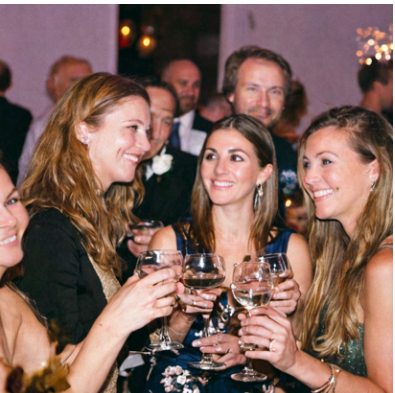
Stable Diffusion

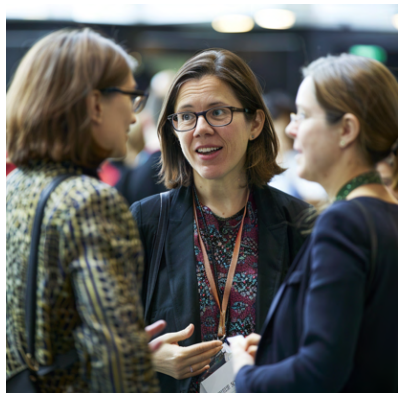
Midjourney

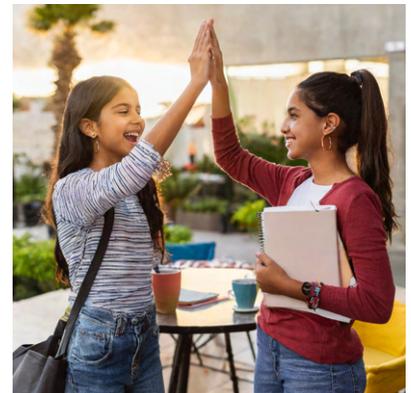
Firefly

## 0.3    Factors that Affect Identifiability of AI Images

The identifiability of an AI-generated image can depend on pose complexity, background detail, the number of people in the image, the size of the faces in the image, and a number of other factors that we will highlight throughout this guide.

For example, AI-generated images of standard photographic poses like a face portrait offer fewer potential cues relative to candid photographs of people interacting with each other. Background detail is important, and when the background is blurred, which can be the natural result of an out of focus lens in portrait photography, there are significantly fewer cues than when the details are present.

On the other hand, when there are many people in an image, the image becomes more complex, resulting in greater possibilities for something in the AI-generation to look unnatural. When there are multiple people in an image, the size of faces are naturally smaller than a face portrait image, and diffusion models often leave artifacts that become obvious when focusing and zooming in on these faces.

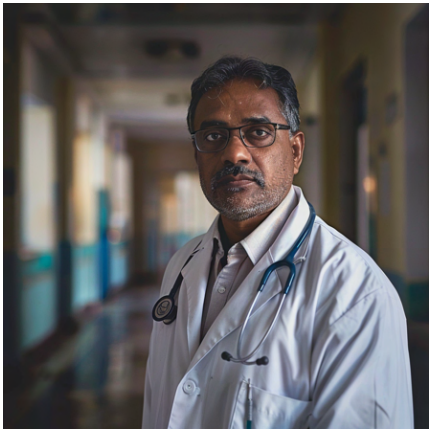
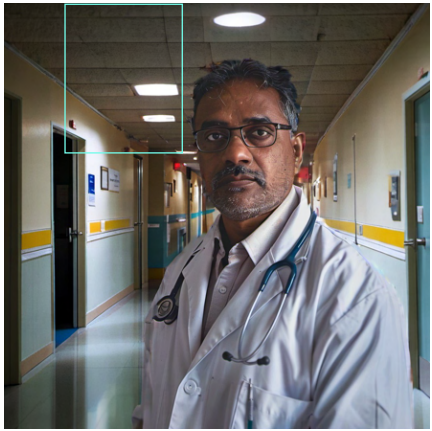

Forcing background detail in the smooth blurred background of the first image results in a less convincing texture and creates visible artifacts in the details such as the irregular ceiling tiles and lighting placement.

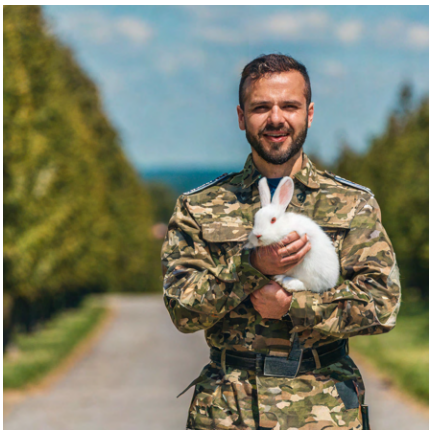
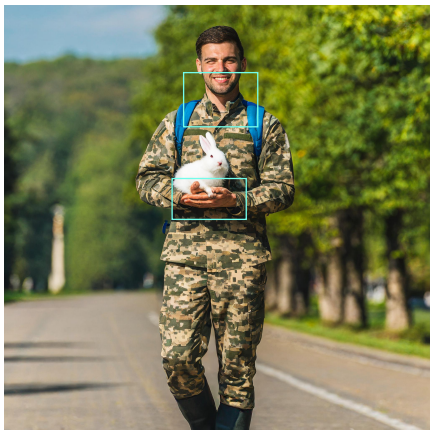
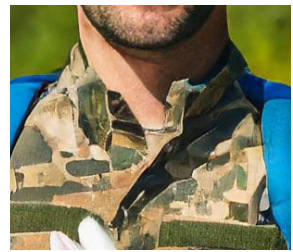
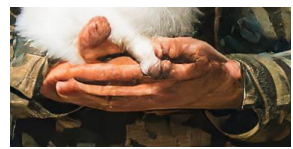

Smaller details in an image allow for more opportunities for error. The portrait image has no obvious artifacts, while the full-body image has obvious errors in the hands and around the collar.

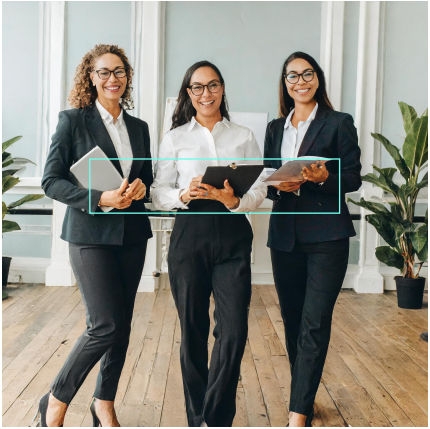

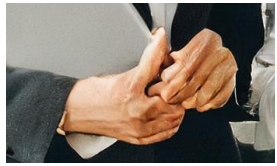 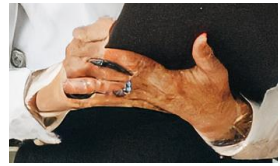 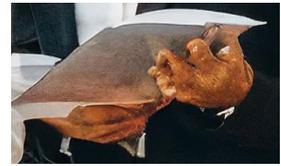

More people in the image allow more opportunities for errors.

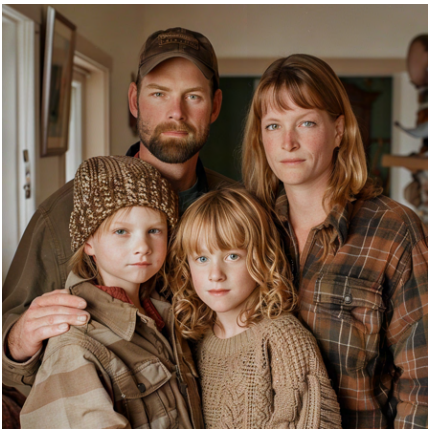

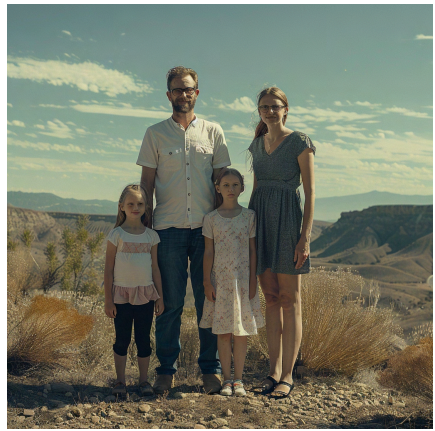

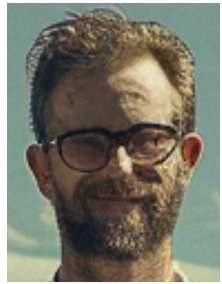 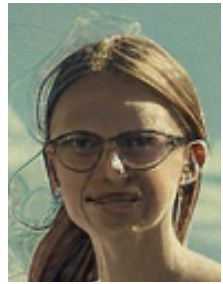 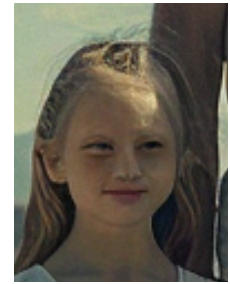 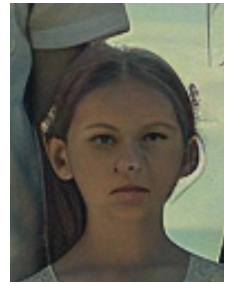

Both of these images are family portraits of an American family. However, when looking closely at the faces of the image on the right where the people are smaller in the image, it becomes obvious that their faces are significantly under-generated in comparison.

**Introduction**

Another significant factor that impacts the identifiability of AI-generated images is the resolution of the image. Low quality and low resolution images reduce the information available to assess the image. Compression artifacts may also distort parts of the image.

Original Resolution

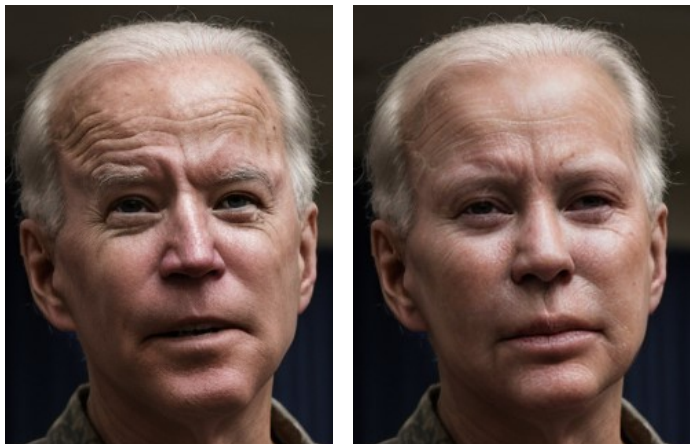

In these two different AI-generated images of Joe Biden, one looks realistic and the other looks slightly like his Republican rival Donald Trump. These images were generated by a custom trained Stable Diffusion model using real images of Joe Biden and Donald Trump.

1/4 Resolution

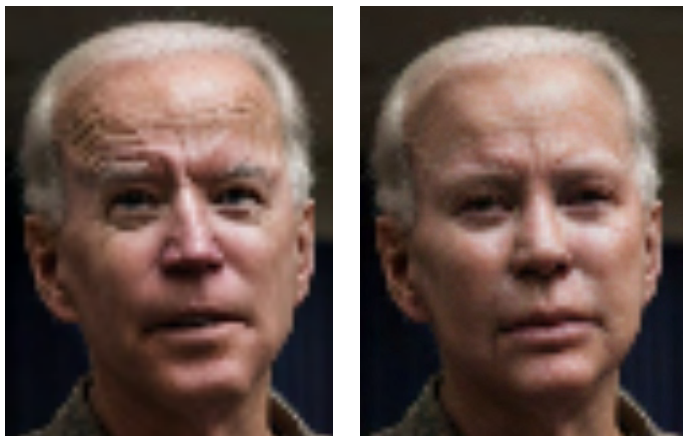

1/8 Resolution

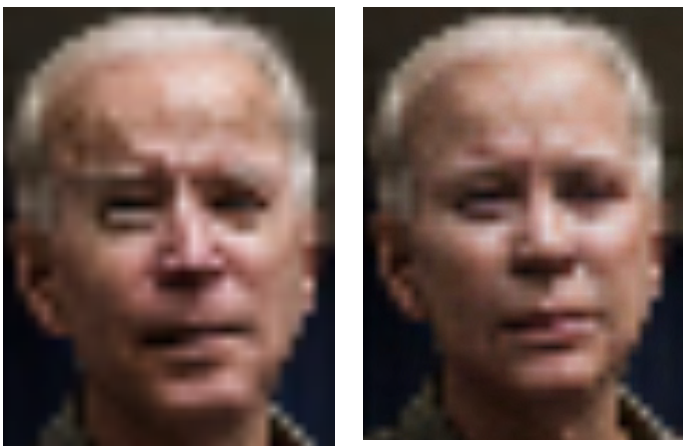

At 1/8 resolution, the features that distinguish the two images are difficult to observe and they appear a lot more like the same person.

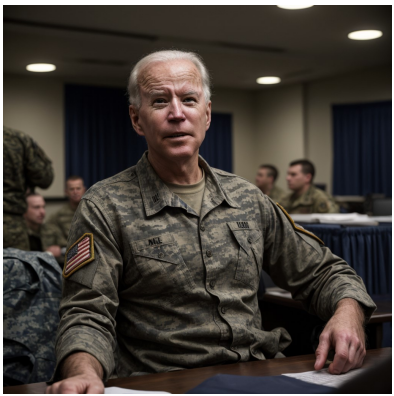

AI-generated Biden image shown in the left column.

AI-generated Biden x Trump face image shown in the right column.

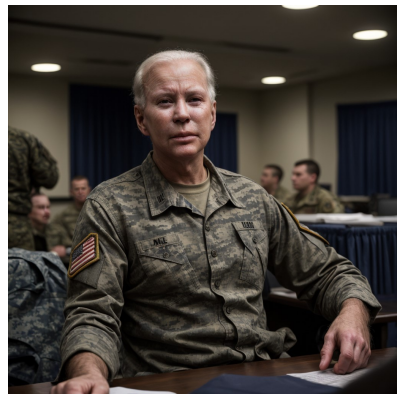

**Introduction**

While there are computer vision models that can identify statistical regularities in the patterns of pixels in AI-generated images, in this guide we discuss semantically meaningful artifacts and implausibilities. Simple edits to AI-generated image such as cropping, color editing, and downgrading mislead state-of-the-art detection models [3]. By bringing attention to visible artifacts in this guide, we aim to enhance humans' abilities to robustly detect AI-generated images and avoid the brittleness that often afflicts AI detection models.

## How to read this guide:

We have curated both AI-generated and real images to visually demonstrate the kinds of artifacts and implausibilities that may help you tell the difference between the two. However, the goal of this guide is to help you to develop a nuanced understanding of when and how visible artifacts may or may not be used to identify synthetic content. We will sometimes show real images that appear to have visible artifacts and AI-generated images that do not have readily identifiable visible artifacts. As this can get confusing, we show real images in a blue box and AI-generated images without obvious visible artifacts in an orange box.

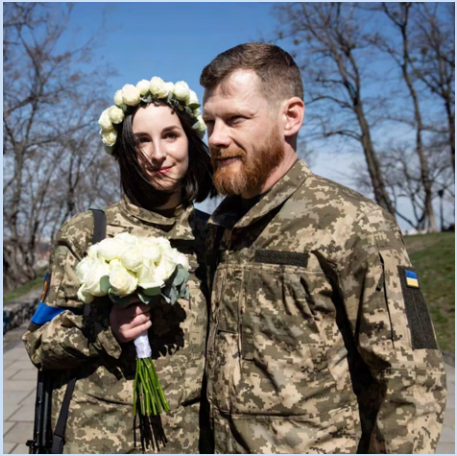

The Ukrainian Soldier From New York's March Cover Got Married. Photo by Mikhail Palinchak. (New Yorker)

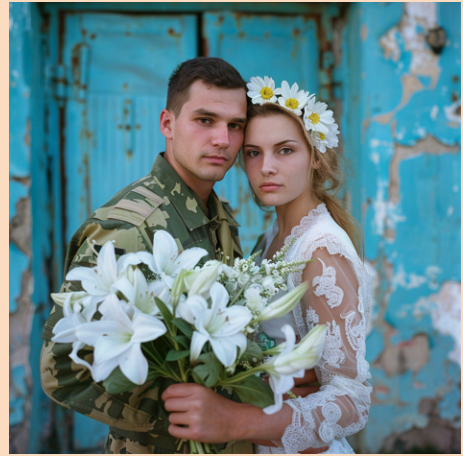

Image made in Midjourney with the prompt: "ukrainian family photo ukrainian soldier and his wife who is a bride carrying white flowers and wearing white flowers in her head high resolution and realistic outside in daylight, street photography style"

# 01.

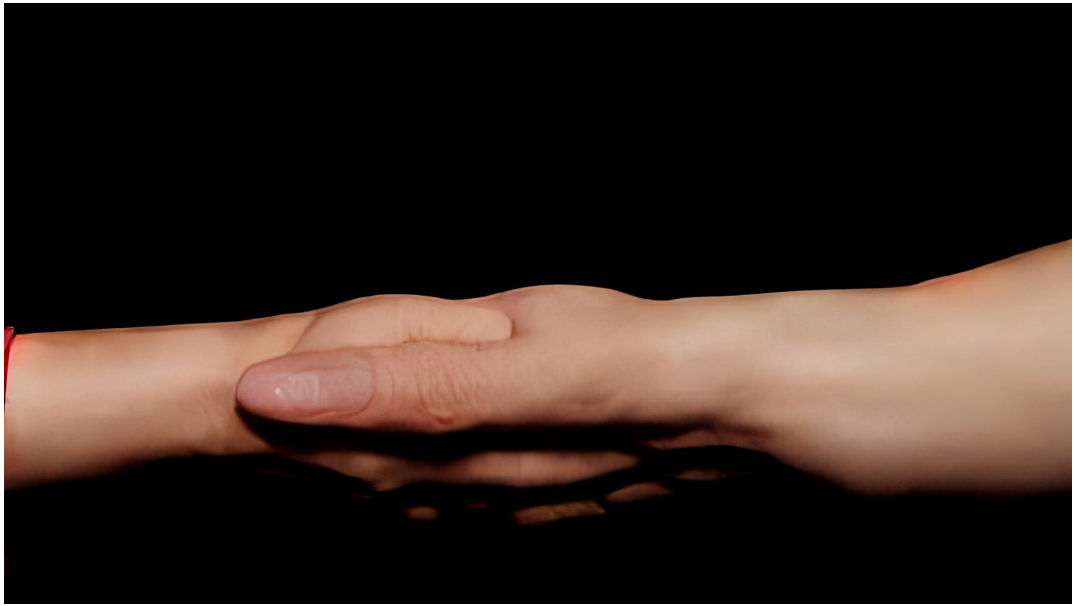

# Anatomical Implausibilities

Anatomical implausibilities can be useful for identifying images likely to be generated by AI. While they do not always appear in AI-generated images of people, nor do anatomical implausibilities necessarily indicate that an image is AI-generated, we will take a look at some artifacts of hands, eyes, teeth, and bodies that appear disproportionately in AI-generated images. Importantly, all bodies are different and what is considered anatomically implausible is meant to guide attention towards signals for identifying AI-generated images.

<table>
<tr><td>1.1</td><td><strong>Hands</strong></td></tr>
<tr><td>1.2</td><td><strong>Eyes</strong></td></tr>
<tr><td>1.3</td><td><strong>Teeth</strong></td></tr>
<tr><td>1.4</td><td><strong>Bodies</strong></td></tr>
<tr><td>1.5</td><td><strong>Merged Bodies</strong></td></tr>
<tr><td>1.6</td><td><strong>Biometric Artifacts</strong></td></tr>
</table>

## 1.1   Hands

Take a close look at the hands in the following portrait images. People in AI-generated images often have missing fingers, extra fingers, merged fingers, nonexistent fingernails, and unlikely hand proportions.

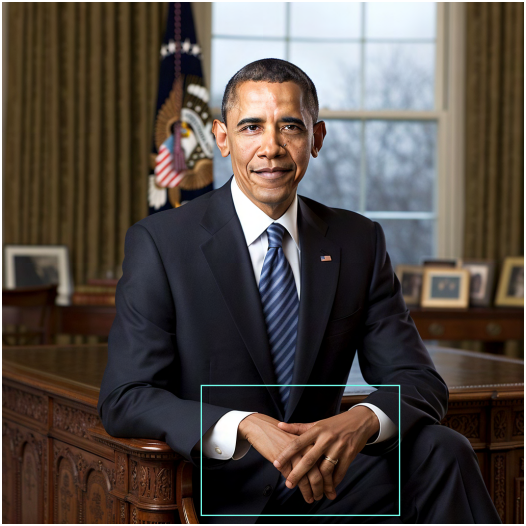

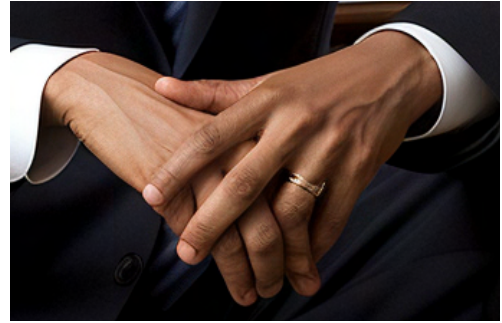

Obama has two hands and not one merged together ...

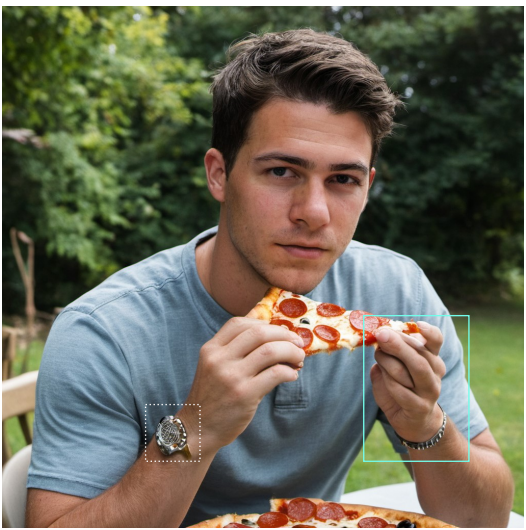

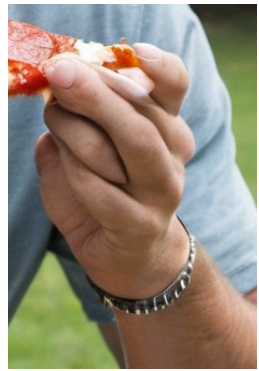

Don't we all dream of having some extra fingers to hold a slice of pizza?

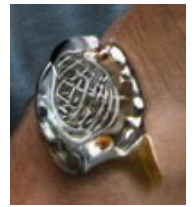

Images often have multiple implausibilities. We'll discuss the watch later in the Functional Implausibilities section.

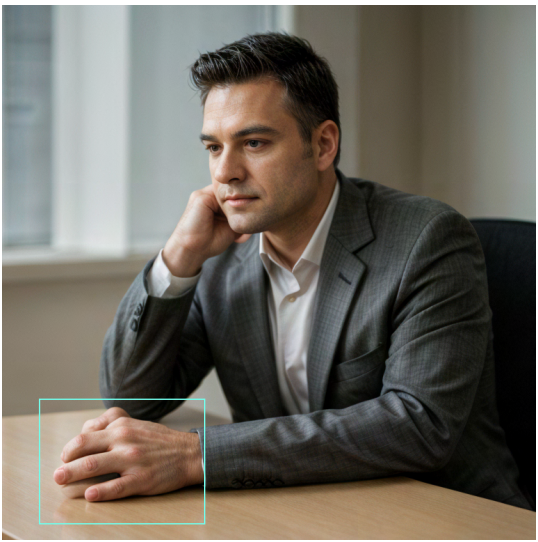

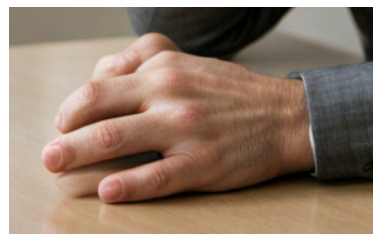

He may be missing a finger here or his thumb appears detached.

## 1.2 Eyes

People in AI-generated images may have misaligned pupils, pupils that aren't circular, unnaturally glossy eyes, or what may feel like an empty gaze.

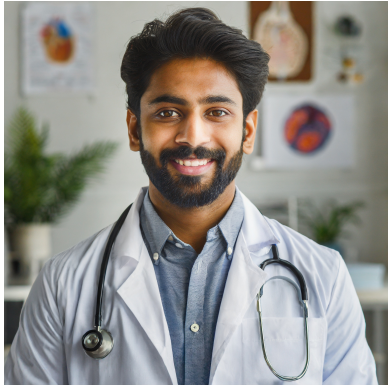 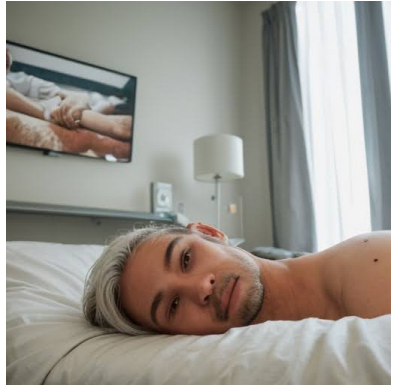 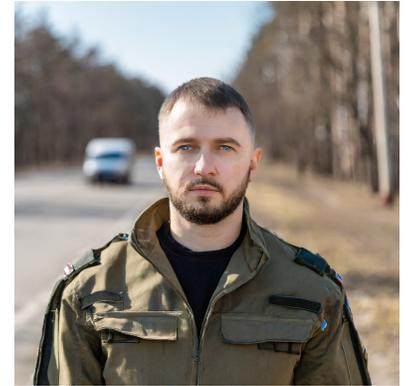

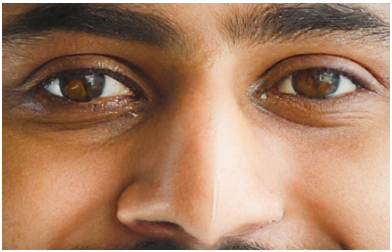 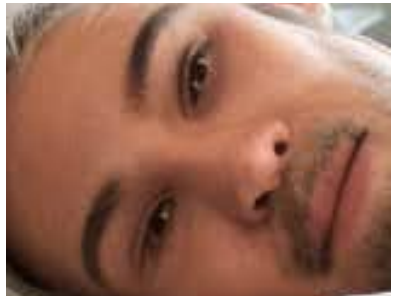 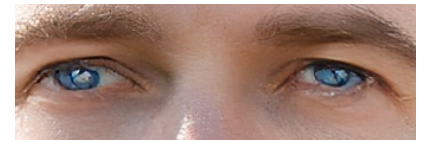

Ah, a comforting hollow gaze from your AI-generated doctor. His eyes look faded and his irises are barely visible.

While you may *feel* like this in the morning, don't worry, you probably don't actually *look* like this. The details of AI-generated eyes can be underdeveloped.

The reflection looks a bit too shiny?

This is a real photograph of Joe Biden. His eyes appear blacked out and hollow, but that is an artifact of the resolution of the image and not AI image generation. Artifacts can often be very subtle and the appearance of one does not necessarily mean it is AI-generated.

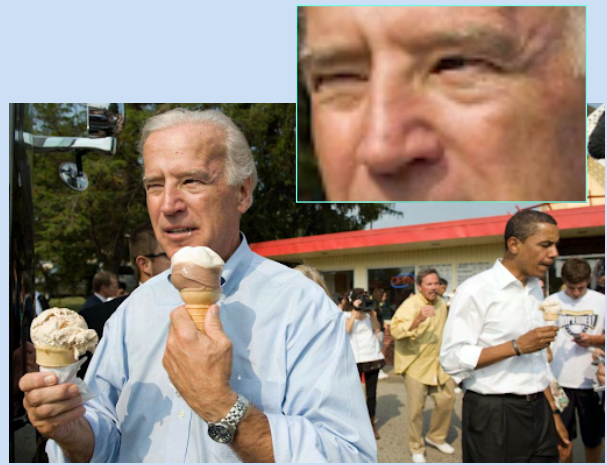

Joe Biden and Barack Obama eating ice cream at the Windmill Ice Cream Shop in Aliquippa, Pennsylvania on Aug. 29, 2008, via *National Ice Cream Day: Joe Biden's Ice Cream Obsession* (TIME).

**01. Anatomical Implausibilities**

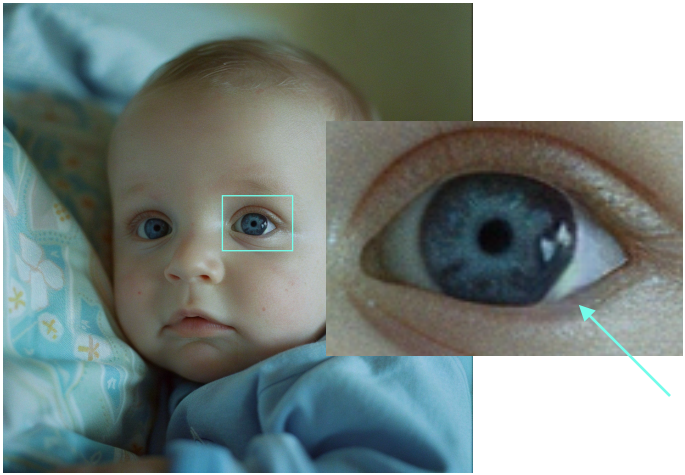

This baby's pupil is distorted around the bottom right edge.

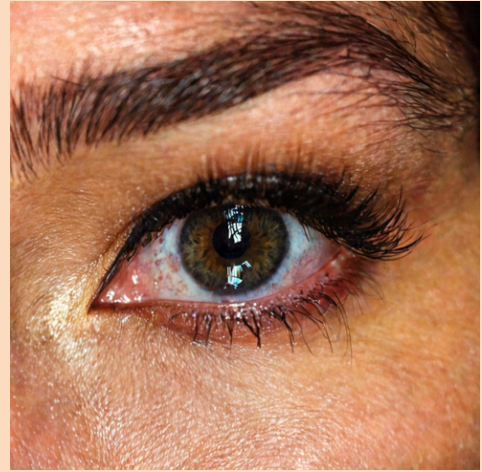

This is an AI-generated image of a zoomed in eye with no obvious visible artifacts.

Glossy eyes that look bright may not always be AI-generated. Fashion photography often features eyes shot in specific lighting conditions or edited to be very clear.

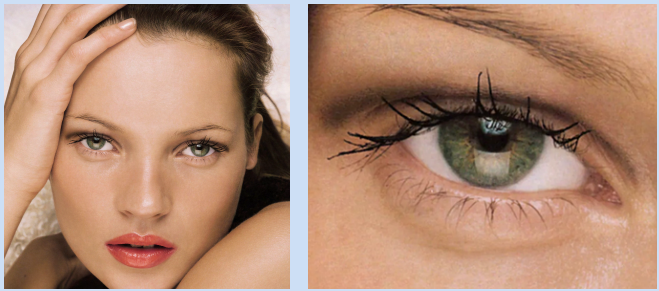

Kate Moss by Nick Knight 1988 (The Guardian)

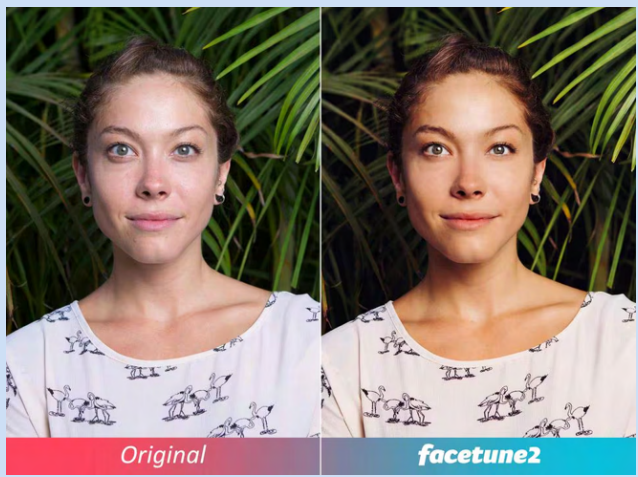

Teenagers have long used Facetune to edit themselves, but what about when it comes to the rest of us? (Boston Globe)

Easily accessible photo editing apps like FaceTune or filters on Instagram, Snapchat and TikTok also allow people to edit facial features including their skin and eyes. These computational edits may produce similar results, but are distinct from diffusion models that generate a complete image from a prompt.

## 1.3    Teeth

People in AI-generated images may have an unlikely alignment of teeth, numbers of teeth, or an overlapping of teeth and lips.

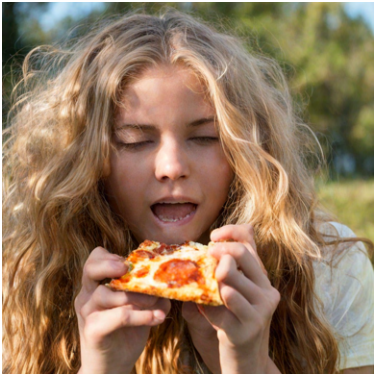 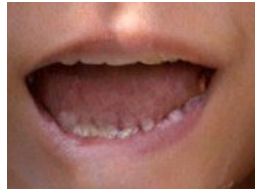

"I can't eat my pizza because my teeth are part of my tongue!"

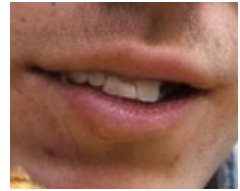

The two front teeth are not proportional and he may be missing some teeth on his left side.

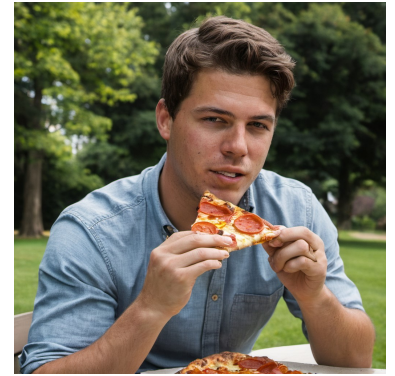

While teeth can be an artifact that signals an AI-generated image, it is also often the case that there are no obvious artifacts related to the teeth and mouth.

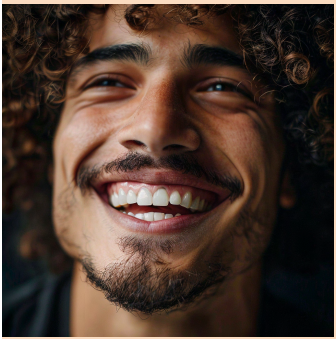 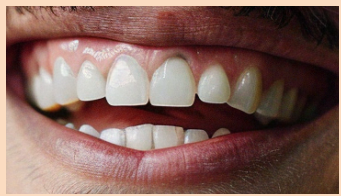 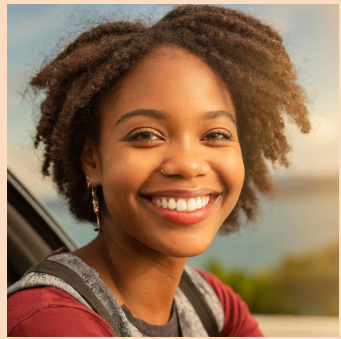 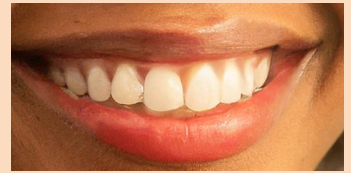

## 1.4    Bodies

People in AI-generated images may have extra limbs, missing limbs, body parts that bend in unlikely ways, and unlikely proportions of body parts.

**01. Anatomical Implausibilities**

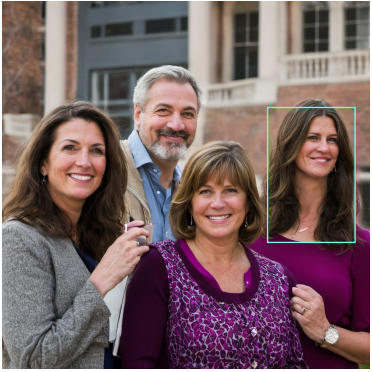
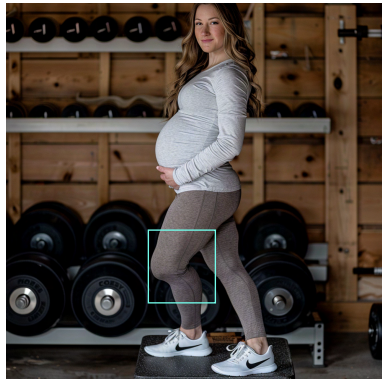
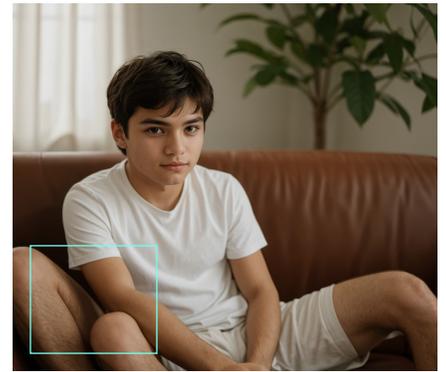

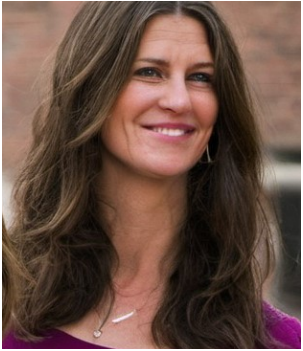
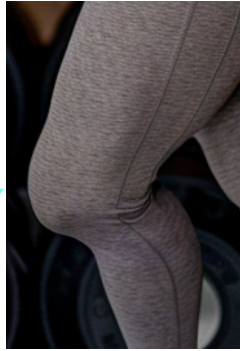
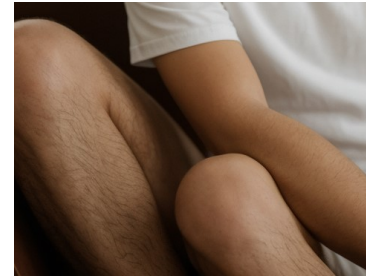

Is her neck a bit long?
A muscular knee cap?
Who's extra leg is that?

## 1.5 Merged Bodies

AI image generators often fail to distinguish between the body parts of different people. This can result in merged body parts.

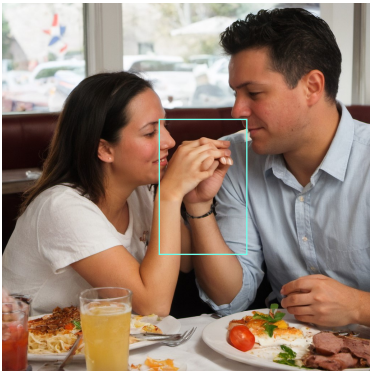
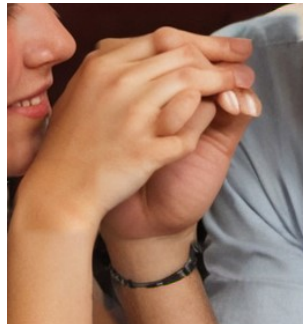

This couple is coupled together. Their fingers merge into one.

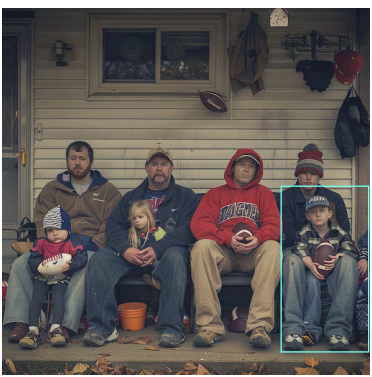
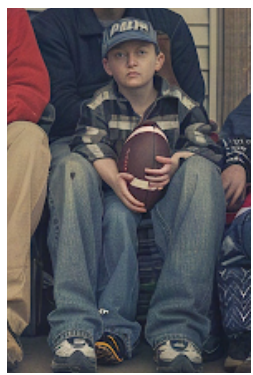

This boy seems to emerge out of the legs of the boy behind him.

## 1.6    Biometric Artifacts

Biometrics are physical characteristics that are unique to individuals and can be used to identify specific people. If an image depicts an individual you have access to other photos of (such as public figures), biometric features such as the size, shape, and proportions of the ears, nose, and mouth can be compared.

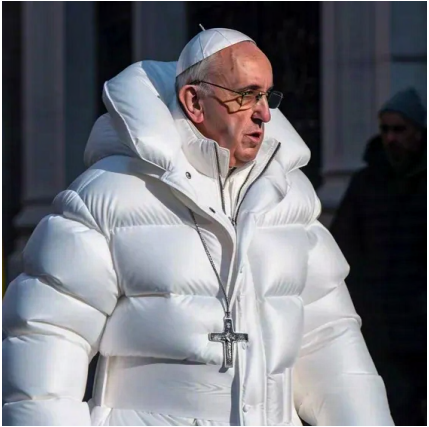 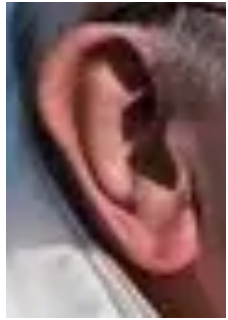 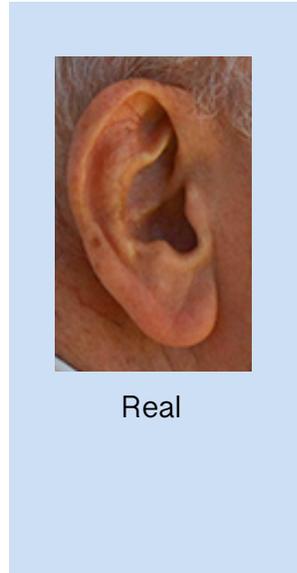 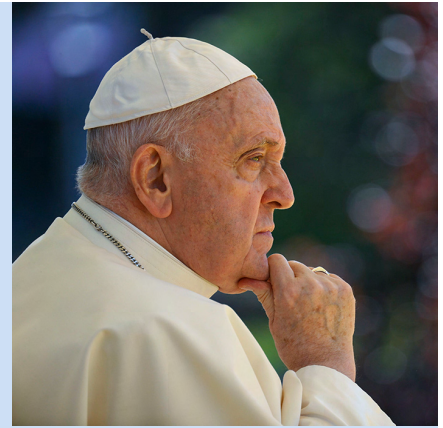

AI

Real

r/midjourney (Reddit)

The center of the pope's earlobes and general contours appear noticeably different in the two images.

Pope Francis at the Portuguese Catholic University in Lisbon, Portugal on Aug. 3, 2023 via Vatican Media (America Magazine)

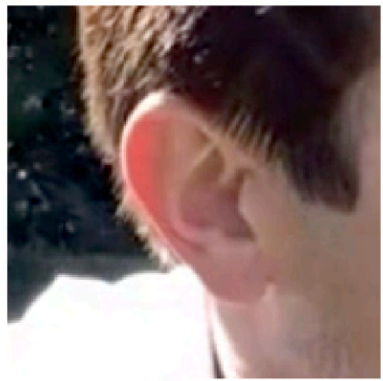 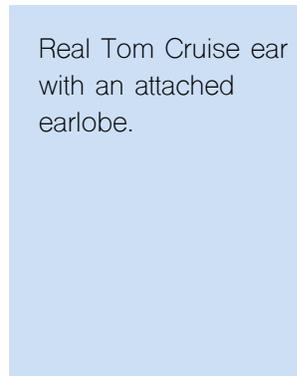 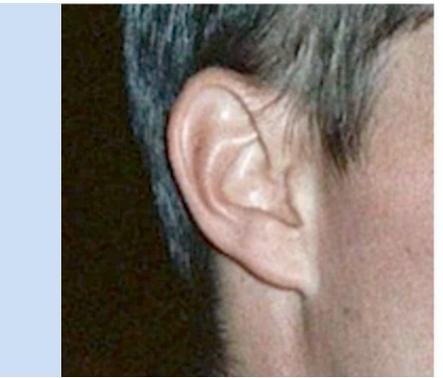

Image of @deeptomcruise from TikTok with a detached earlobe.

Real Tom Cruise ear with an attached earlobe.

Images from Detecting Deep-Fake Videos from Aural and Oral Dynamics, Shruti Agarwal and Hany Farid

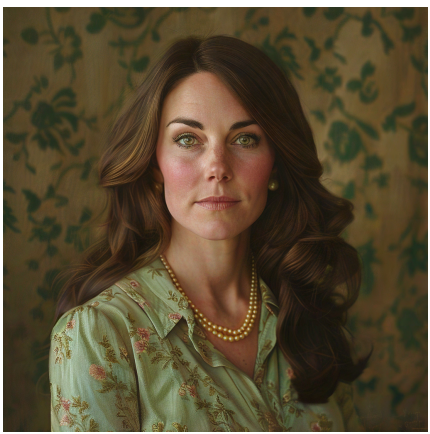 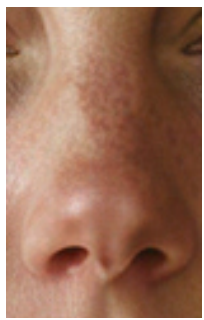 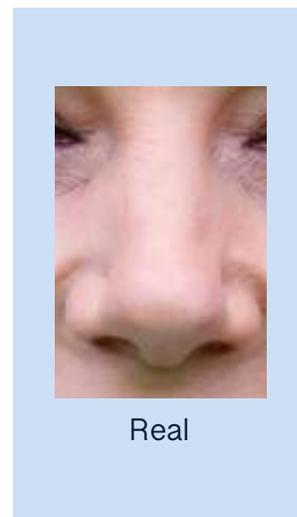 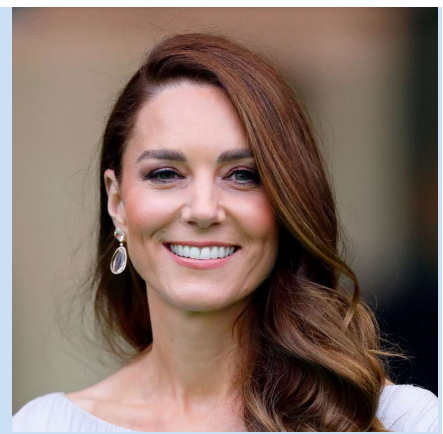

AI

Real

The bridge of Kate Middleton's nose is different in the AI image vs. the real photograph.

Kate Middleton portrait (Max Mumby/ Indigo/Getty Images)

**01. Anatomical Implausibilities**

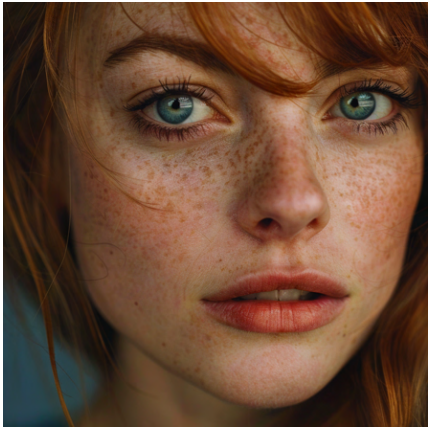

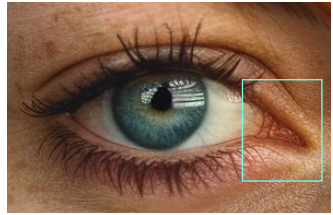

AI

Emma Stone's tear duct is deeper in the AI-generated image in comparison to the real photograph.

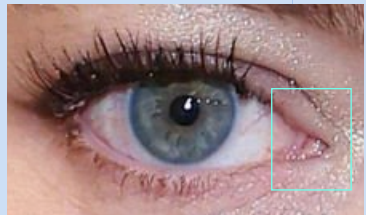

Real

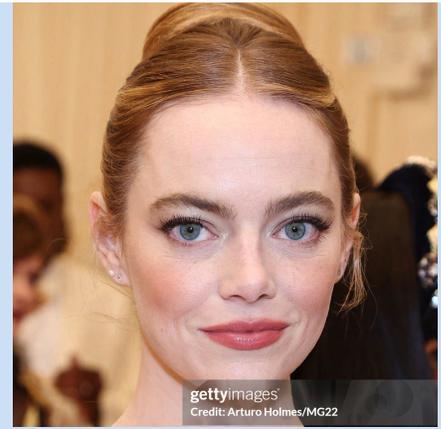

Emma Stone portrait (Getty Images)

## Summary

Anatomical implausibilities in AI-generated images can occur in various body parts and can range from obvious to subtle. If you are assessing an image where the hands are visible, first look at the hands. If you are assessing an image of a group of people, look for merged body parts. Look closely at the limbs of each individual person and see if they disappear behind objects or combine into other people. Also check for any unnaturally empty gazes. If you are looking at a full-body image of an individual, be sure to zoom into the hands and facial features. Artifacts in the eyes and teeth such as overly shiny eyes or overlapping of teeth and mouth may be evident upon a closer look. If an image depicts a known individual with other reference images, try comparing the size, shape, contours, and proportions of specific facial features.

Always keep in mind that there is no strict definition of an anatomical implausibility. Photo editing, makeup techniques, and compression artifacts can all resemble anatomical implausibilities. If you observe a body part that looks unnatural, it could signal an AI-generated image, but it may not necessarily be conclusive evidence so it is important to look for multiple signals.

Guiding Questions:

- Are there any artifacts in the hands?
- Are there any unnatural proportions in the limbs of the people?
- Are there any body parts that merge between different people?
- Does the gaze of any person look unnatural?
- Do you notice anything unnatural about the eyes or mouth/teeth?
- Does the image appear to depict a person you have other images of? If so, are there any noticeable differences when comparing the size, shape, and proportions of biometric features such as the nose, ears, and mouth with other images of this individual?

# 02.

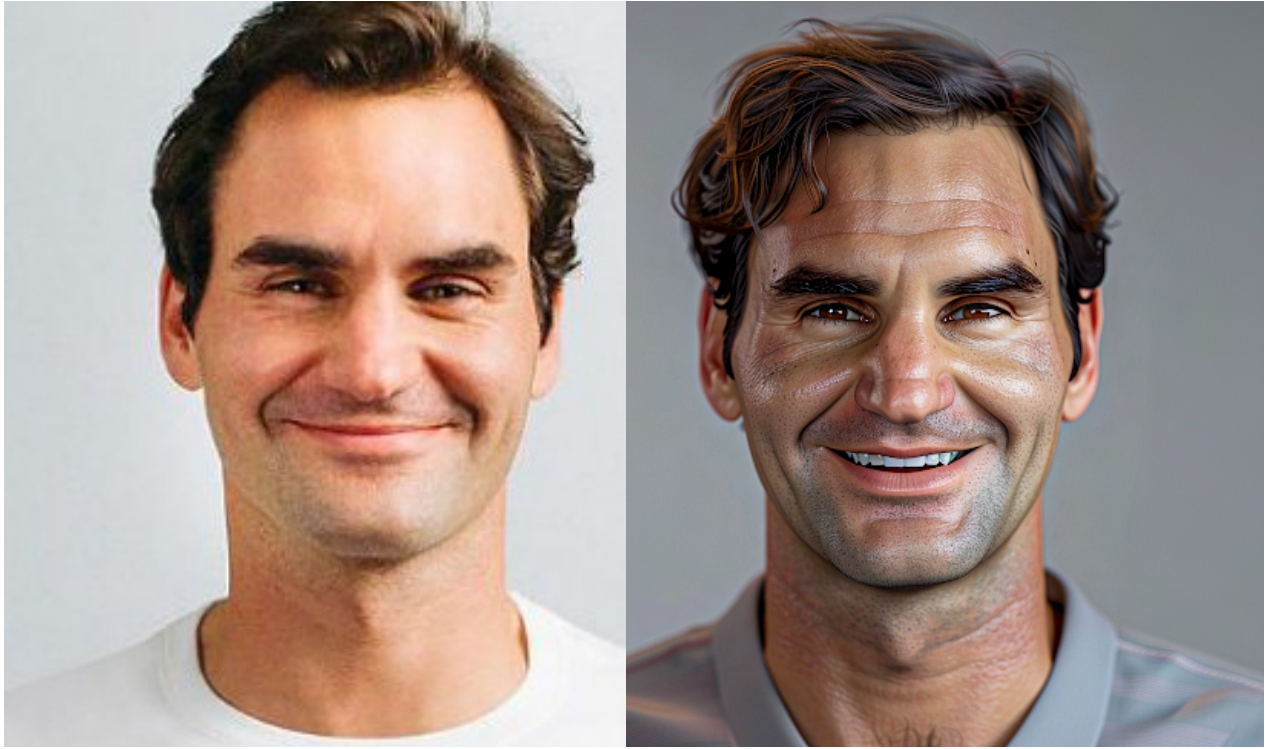

Roger Federer, Aceshowbiz

## Stylistic Artifacts

AI-generated images often look a bit too perfect. The people may have glossy, shiny skin, windswept hair, and they may look like a photo in a magazine or a scene from a movie. Sometimes, parts of an image have a different level of detail or vibrancy compared to the rest. We call these Stylistic Artifacts.

**2.1**    **Plastic Texture**

**2.2**    **Cinematization**

**2.3**    **Hyper-real Detail**

**2.4**    **Inconsistencies in Resolution & Color**

## 2.1  Plastic Texture

Take a look at the AI-generated people below. The texture of their skin can often be characterized as waxy, shiny, cartoonish, or glossy.

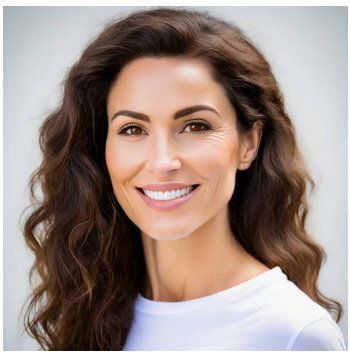 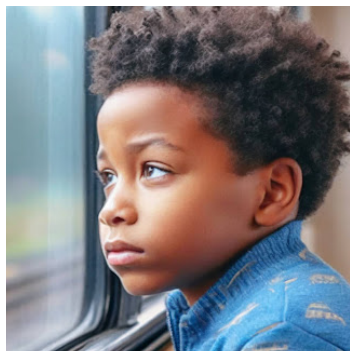 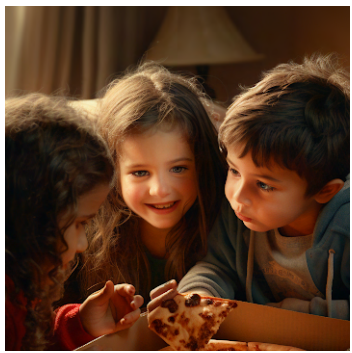 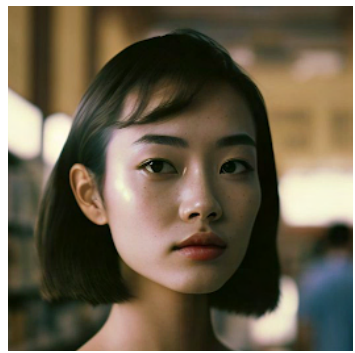

Waxy                Shiny                Cartoonish                Glossy

Given different prompts and settings, the same image can be produced in different styles. Take a look at the two rows below, which present images from two attempts to generate an image of a boy eating pizza at a park in three distinct styles. The first column portrays the boy in the style of a 3D render, the second in a smooth, model photoshoot style, and the third in the style of an iPhone photo. The variation in style was achieved by re-generating an initial image through different combinations of custom-trained Stable Diffusion models.

3D Render                Photoshoot                iPhone Photo

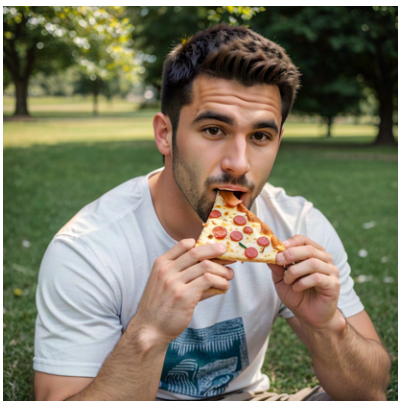 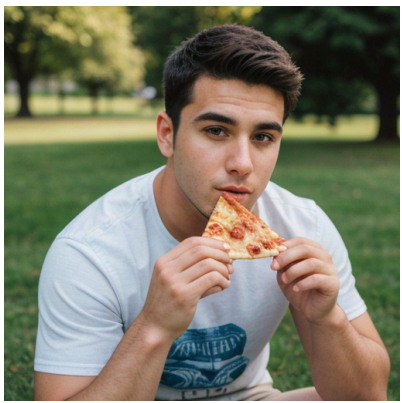 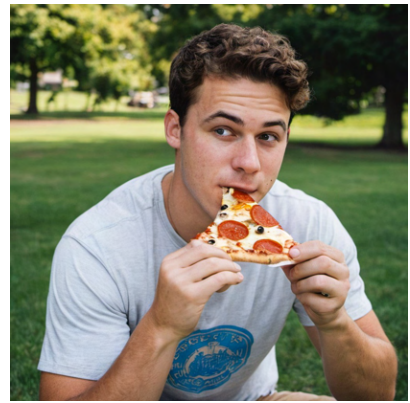

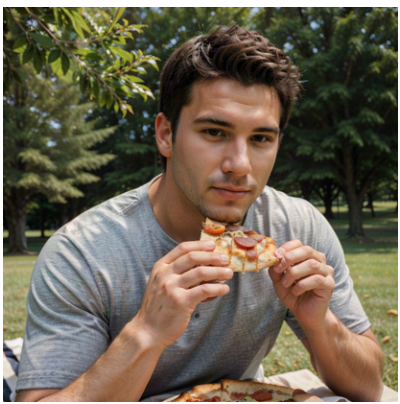 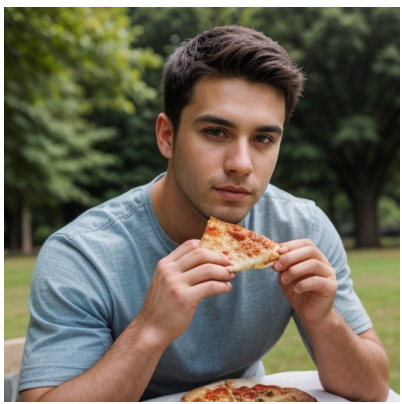 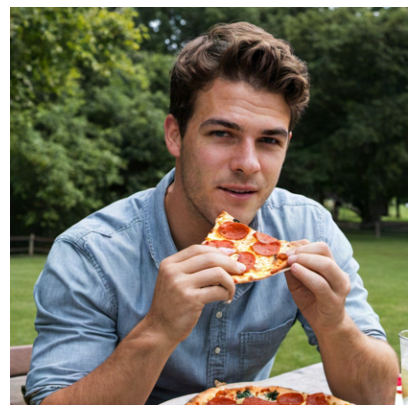

While perfect skin is uncommon in real life, professional photos and fashion photos often portray people with perfect skin.

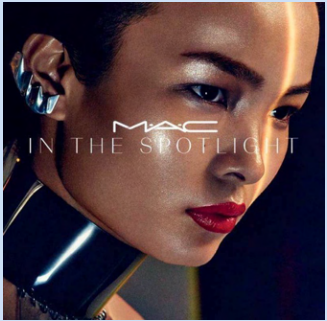

M.A.C Cosmetics S/S 2017–in the spotlight shot by Craig McDean, (models.com)

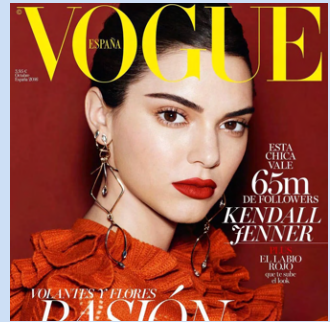

Kendall Jenner (Vogue Spain)

## 2.2    Cinematization

AI-generated images often portray a cinematic or picturesque style that dramatizes the subjects of the image.

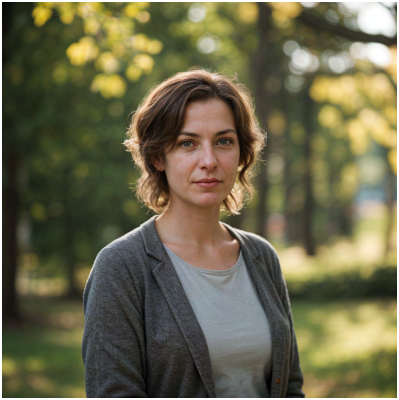
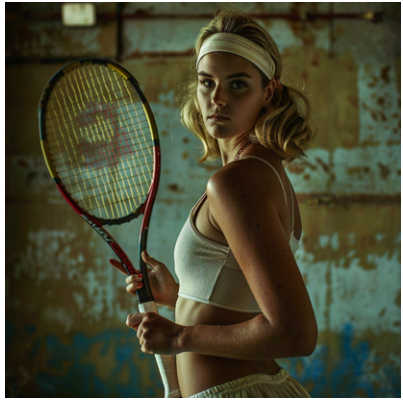
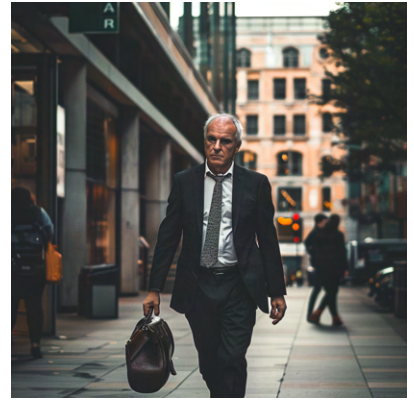

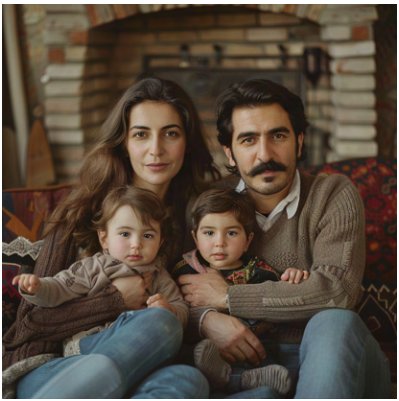
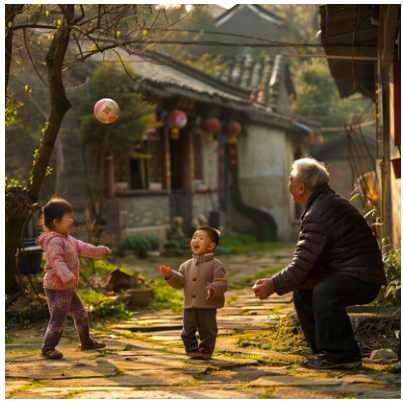
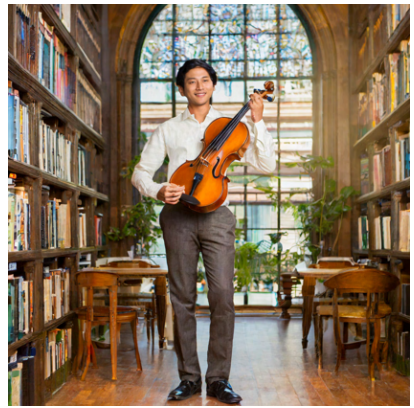

However, it is very common for authentic photographs to feature a cinematic and picturesque style. This can be a product of various color editing procedures on real photographs, the color of the film, or the type of camera used. Photography is an art form that allows for creativity and there is a wide spectrum of styles in real photographs.

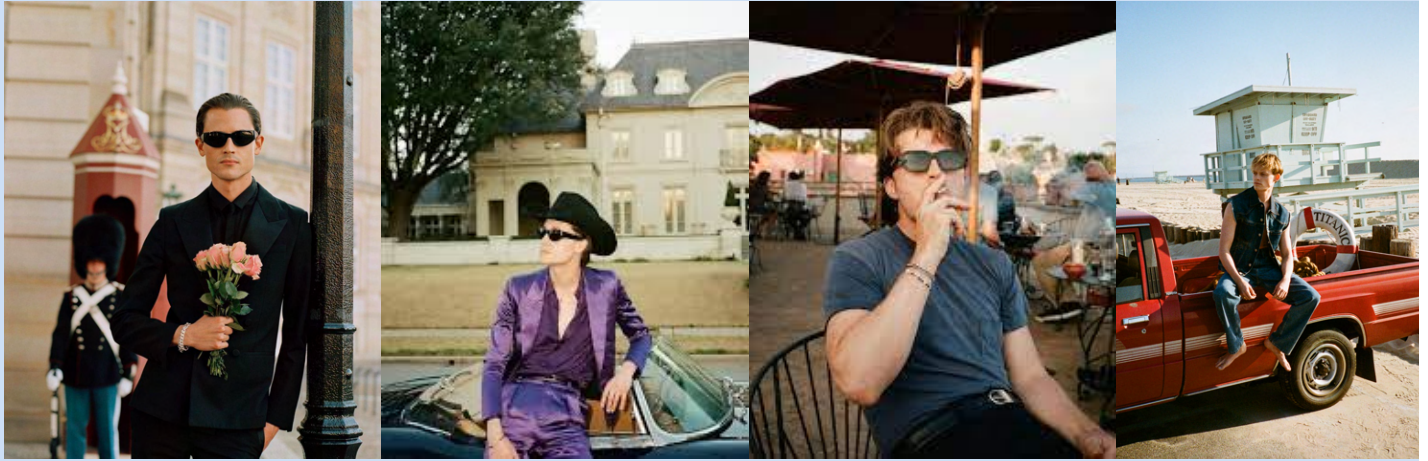

Celebrity portraits by filmmaker and photographer Christian Coppola

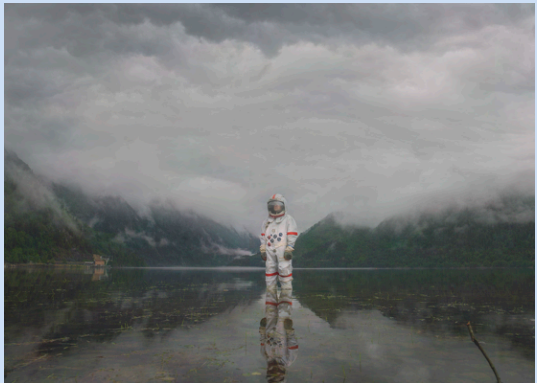

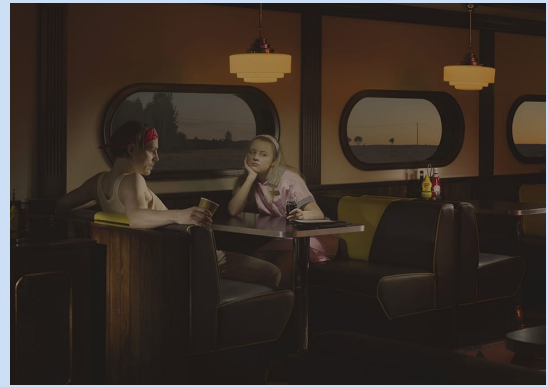

Cinematic photography by Ole Marius Joergensen

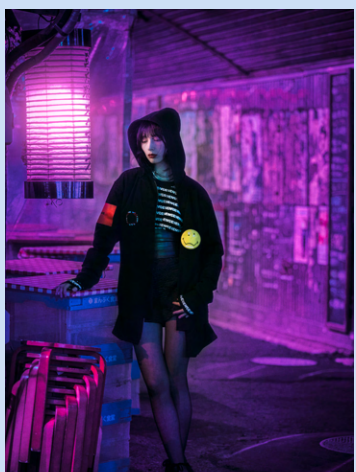

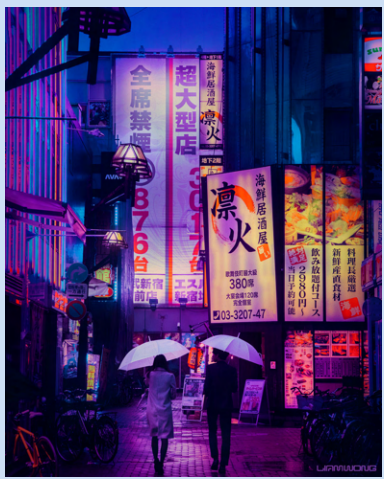

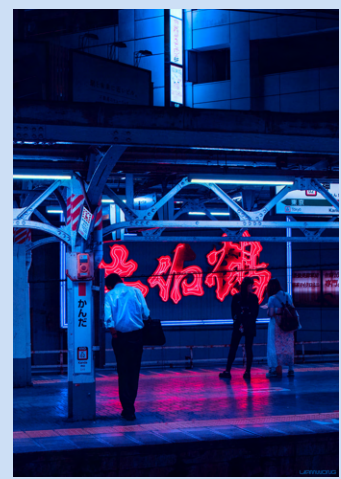

Night photography by Liam Wong

## 2.3    Hyper-Real Detail

AI-generated images can produce an unnatural level of detail in specific parts of an image. This often appears in the way that hair can look excessively soft, fine-grained and windswept in a way that is not consistent with the scene.

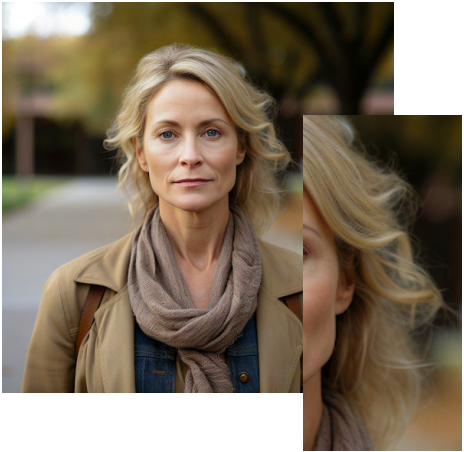

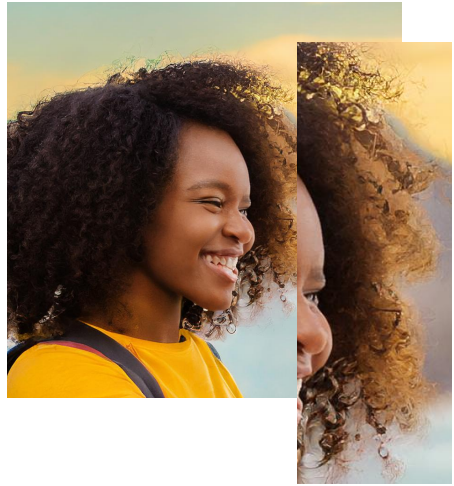

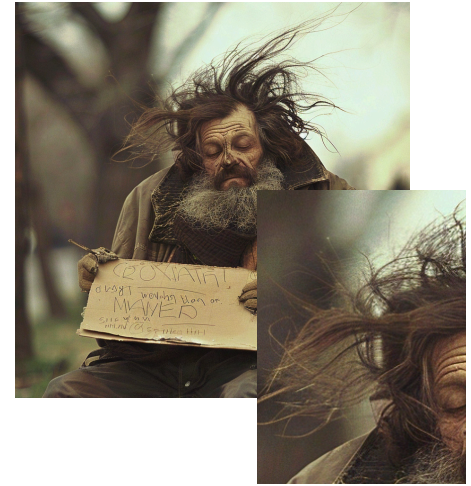

Hair appears wind–swept while all other parts of the image are still.

The color and resolution of the hair is inconsistent in spots.

The homeless man's hair appears unnaturally soft.

Real photographs may portray wind-swept hair as well. Below are three examples of detailed, soft hair blowing in the wind.

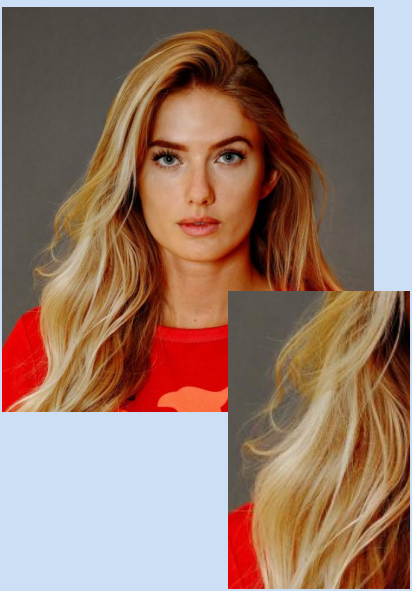

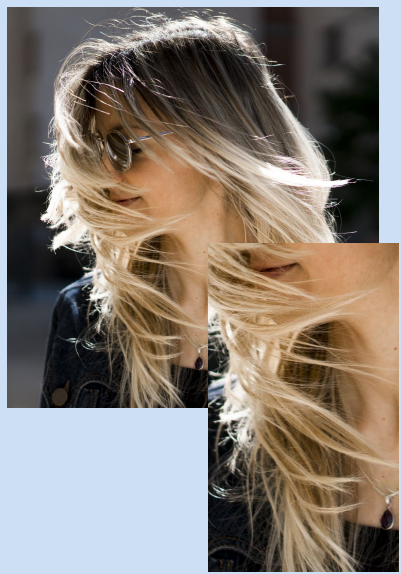

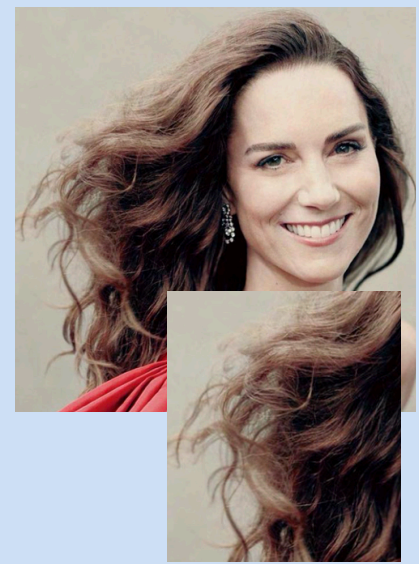

Alicia Schmit Olympics Runner Official Portrait (Welt)

How to have a good hair day everyday, (Red)

Kate Middleton Instagram

## 2.4    Inconsistencies in Resolution & Color

AI-generated images can produce inconsistencies in the style and resolution of different parts of an image. This may appear between the subjects and backgrounds of an image, or in the seams between different objects in an image.

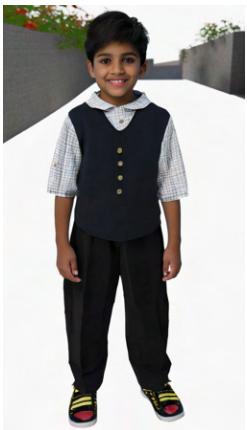

This image looks like an incomplete green screen composition with the subject photoshopped onto an empty white background.

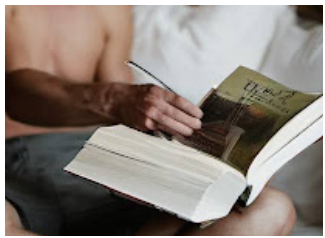

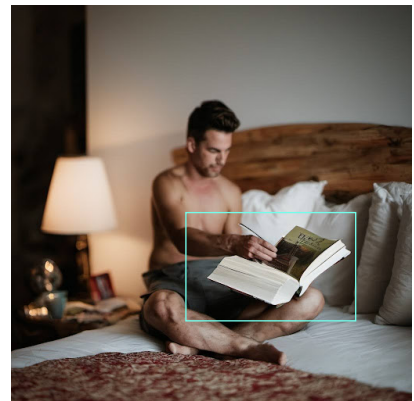

There is a unnatural amount of blur that appears to mask out the hand and book.

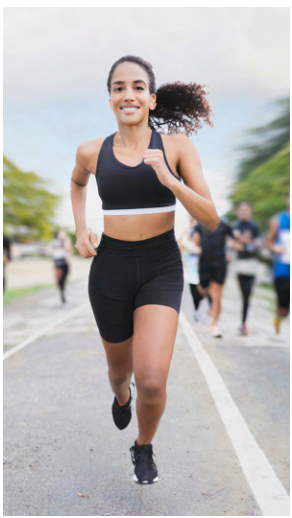

This runner looks like she is photoshopped onto the background from a different image due to the lighting and resolution of the runner in comparison with the rest of the image.

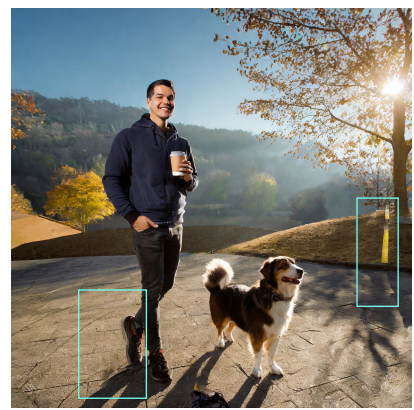

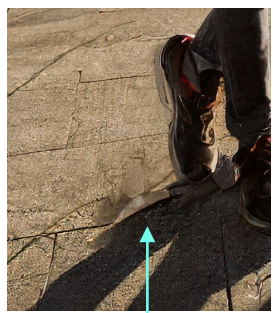

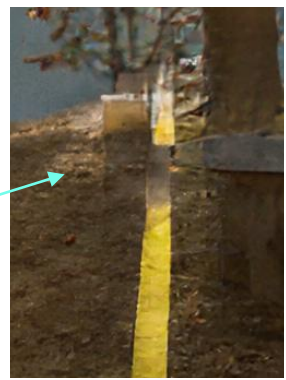

Smudge-like, patchy glitches

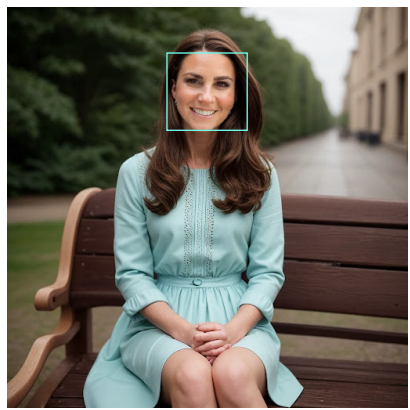

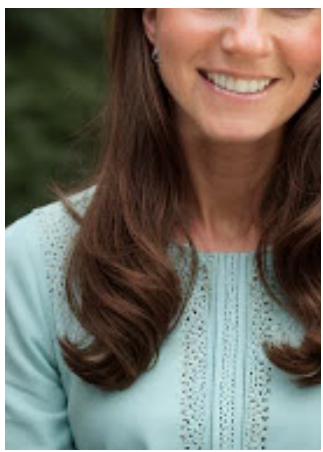

Kate's face looks noticeably brighter than the rest of the image. This kind of mismatch in the face specifically is often seen in face-swapped images.

## Summary

AI-generated images often produce images of people that are waxy, glossy, shiny, and look a bit too perfect. AI-generated images are also often noticeably cinematic and picturesque. These qualities can be very obvious when looking at an image. However, keep in mind that professional photographs are often shot and edited to look clean and cinematic in similar ways, so seeing these features does not necessarily determine if an image was generated by AI. Additionally, look out for inconsistencies in resolution between different subjects or parts of an image as they could signal an AI-generated image, but be conscious of traditional photo editing techniques that can also produce similar looking artifacts.

Guiding Questions:

- Does the person in the image look waxy, glossy, shiny, or plastic-like?
- Does the scene look unnaturally dramatic and cinematic?
- Are there any missing backgrounds or unnatural backgrounds?
- Do different parts of the image look like they are cut out from different scenes?
- Does the face look like it is under different lighting than the rest of the image?
- Are there any smudge-like glitches at the edges of different components in an image?

# 03.

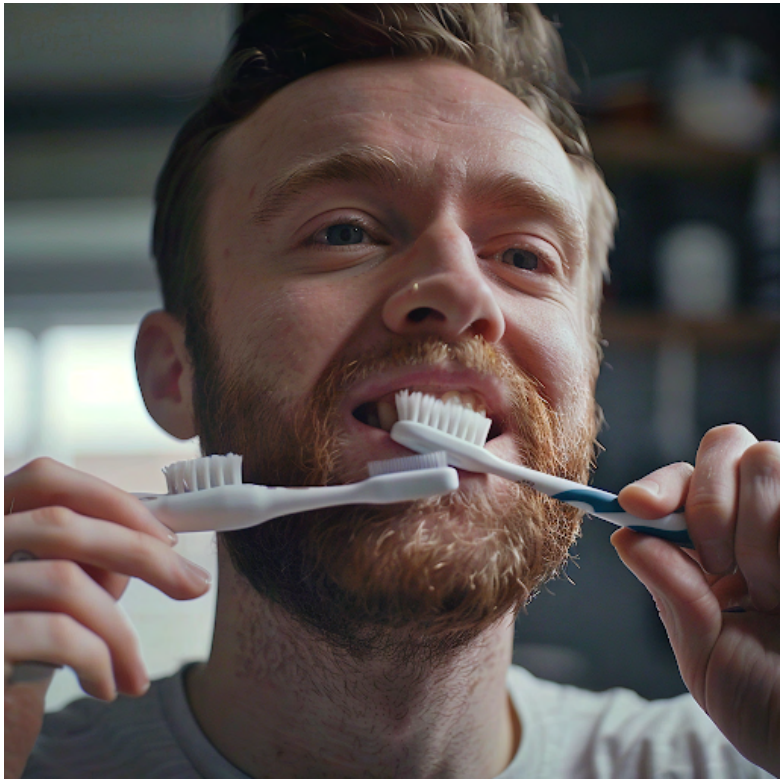

## Functional Implausibilities

AI image generators do not have a structural understanding of how objects in the world work and interact with each other. This can cause a range of functional implausibilities like compositional implausibilities, dysfunctional objects, and atypical designs. Looking closely may also reveal distorted, unresolved details. Incomprehensible text and logos as well as overrepresented features are also a distinct functional implausibility generated by AI.

## 3.1    Compositional Implausibilities

Compositional implausibilities are relations between objects and people that do not conform to real-world mechanical principles. These can include floating or unsupported objects and objects merging into other objects.

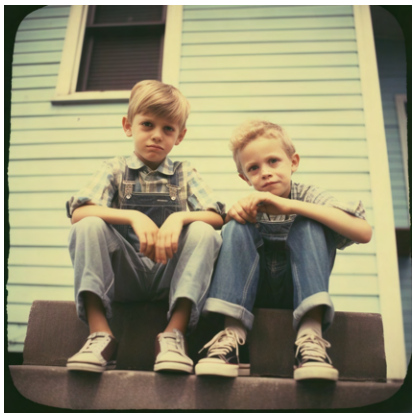

Stairs emerging from side of the building.

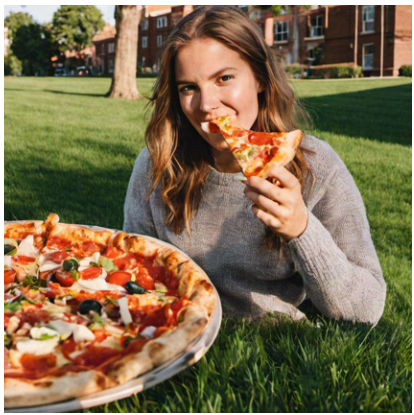

Her torso is growing from the ground.

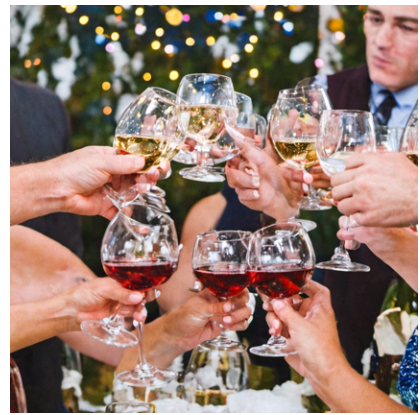

Can you count the number of wine glasses and hands?

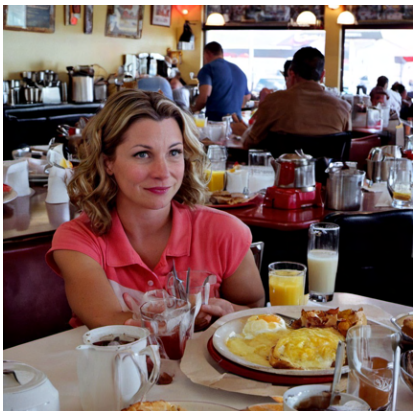

Photo or postmodern montage of diner elements? The details of this image are not quite resolved.

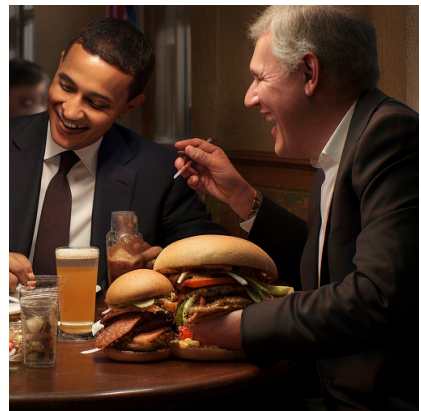

Not quite how we eat a burger.

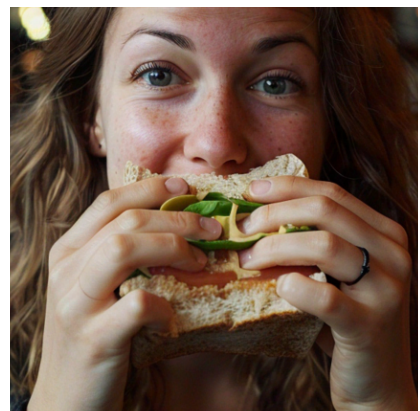

Not quite how we hold a sandwich.

## 3.2    Dysfunctional Objects

AI-generated structures and objects may sometimes appear in a modified form that does not make logical sense, or makes them unusable.

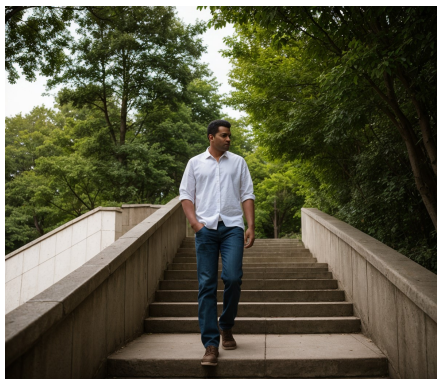

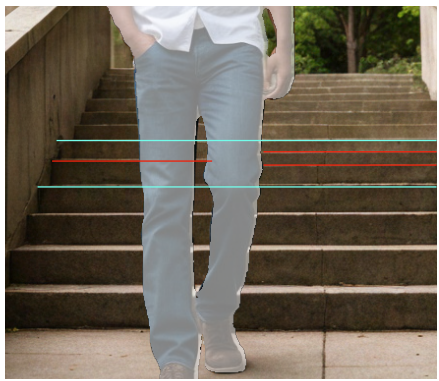

This staircase has a different number of steps to the left and right sides of the man.

## 03. Functional Implausibilities

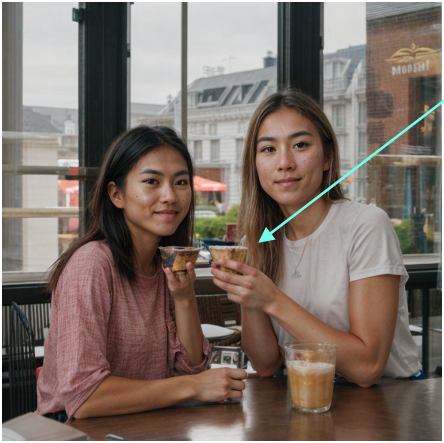

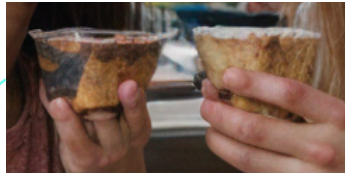

Vaguely coffee and pastry looking cups?

This watch emerges from the arm. It also shows time in lines.

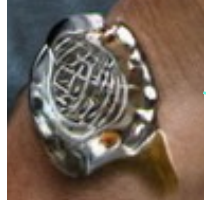

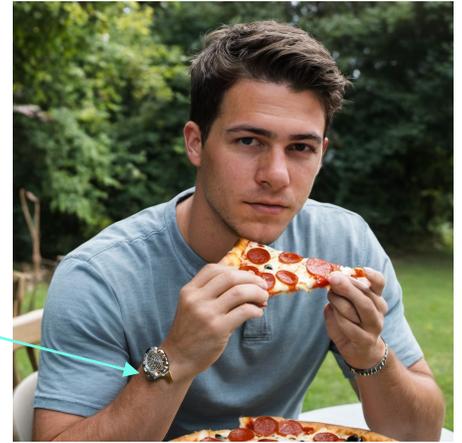

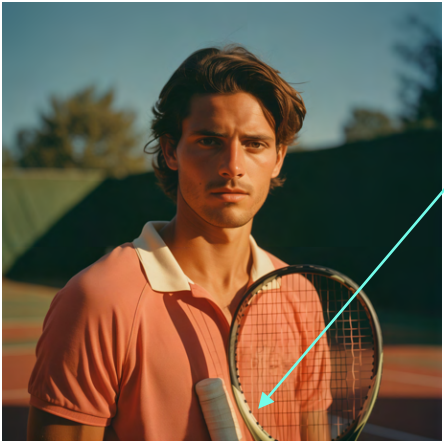

Racket handle appears detached.

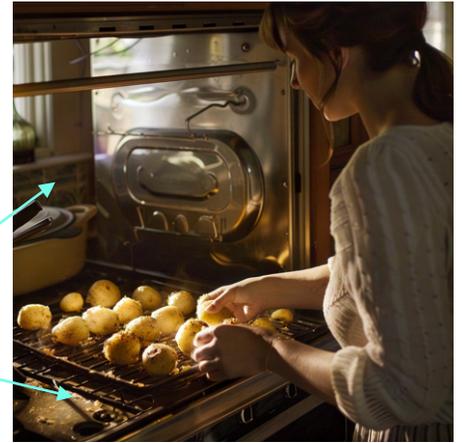

The inside of the oven and stove top are mixed

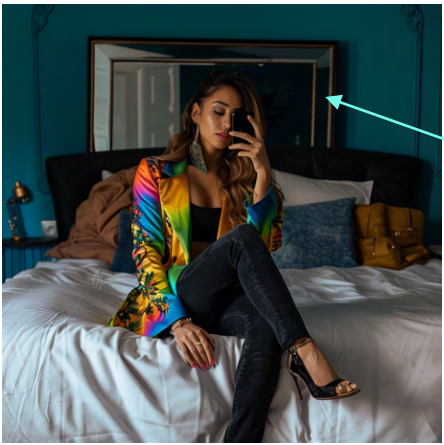

The mirror is placed in a way that defies practicality (it is blocked by a bed)

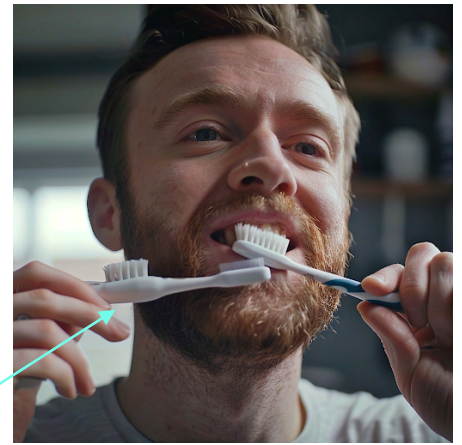

Extra bristles on toothbrush

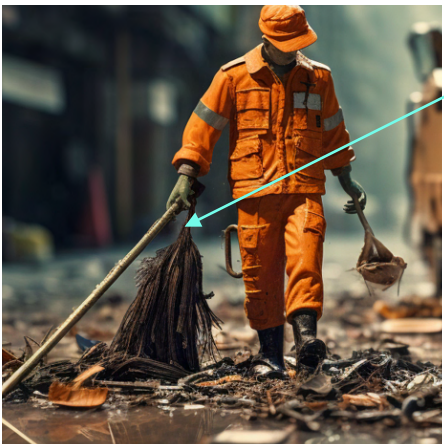

The grip on the broom is broken and unusable.

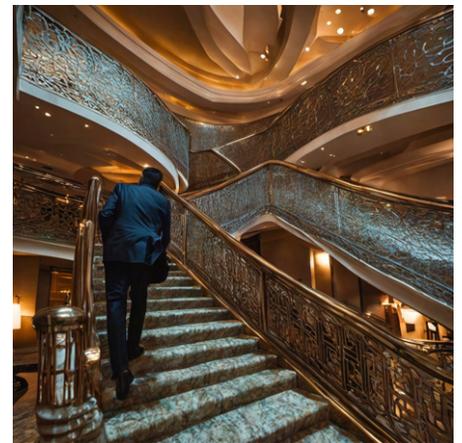

Where does this staircase lead to...?

## 3.3    Detail Rendering

High resolution details are often difficult for AI image generators. Glitches can reveal themselves when you zoom into the details of objects.

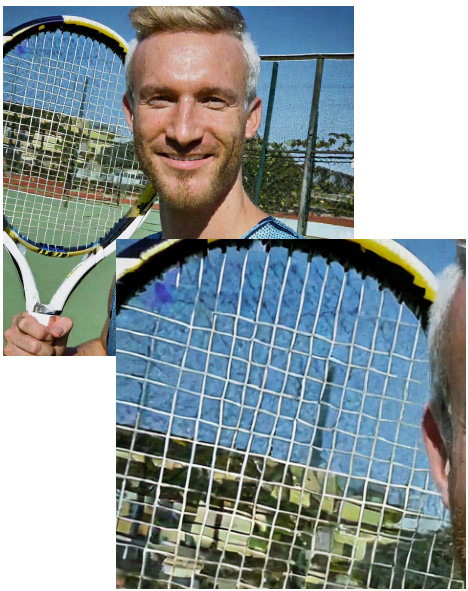

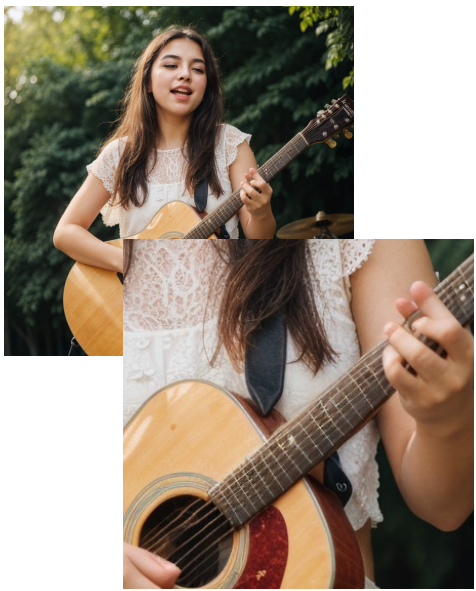

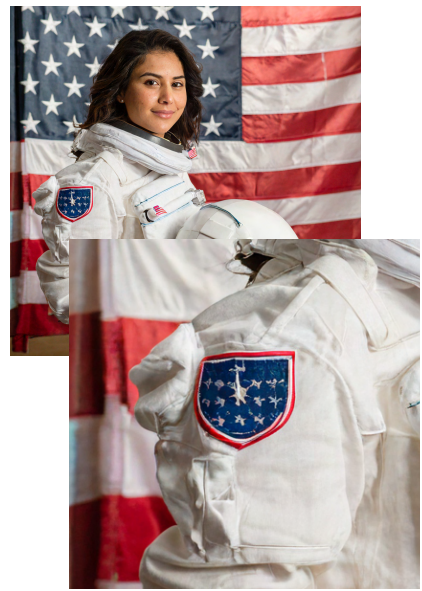

The strings of the tennis racket strings appear misaligned.

There is a varying number of strings across the fretboard of the guitar and the strings appear loose.

Small details in clothing often look glitched and distorted.

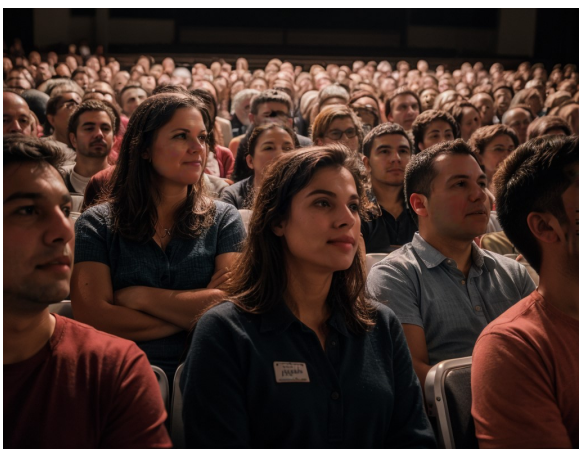

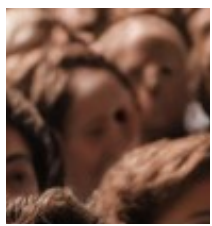

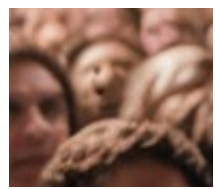

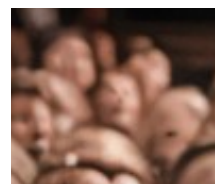

If you zoom into some of the small faces in the back of this crowd, it can get a bit uncanny.

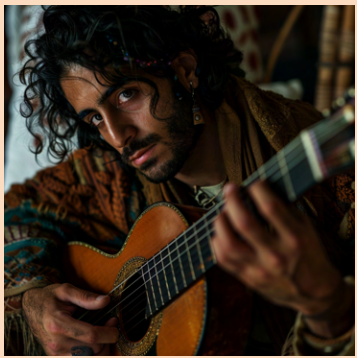

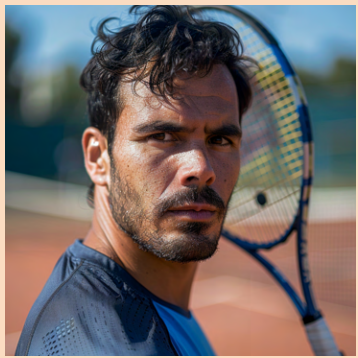

Features that we point out as artifacts now may not be visible as the technology evolves. AI-generated images can include strings on sports equipment and musical instruments that look realistic and artifacts in small details can be obscured through blurring.

**03. Functional Implausibilities**

Object detail may not necessarily be wrong. Sometimes AI images will produce objects with atypical designs. This is often seen in clothing.

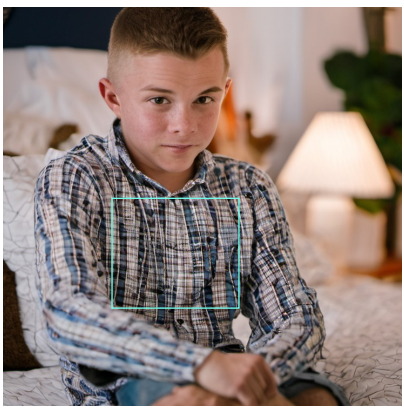

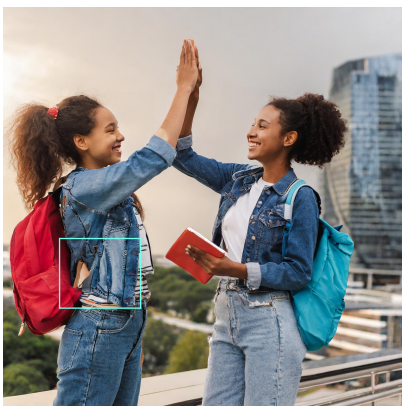

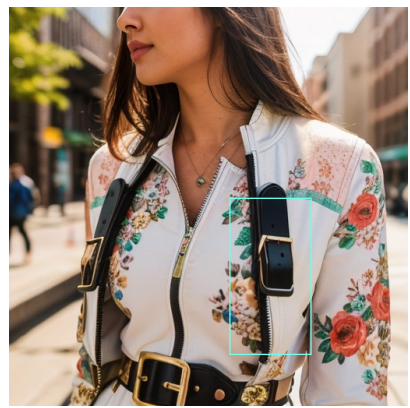

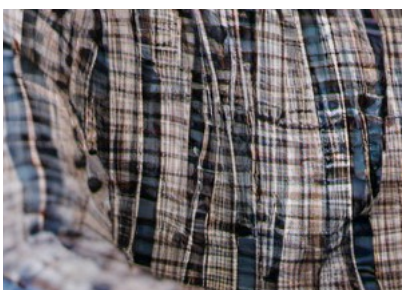

The lines are not straight in this irregular plaid pattern.

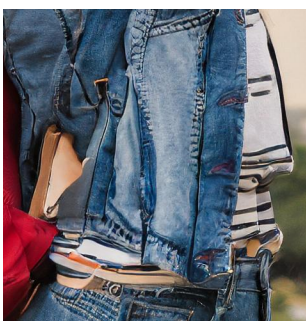

Atypical stitching on the denim jacket and irregular stripes on the shirt. The backpack strap also becomes a part of the shirt.

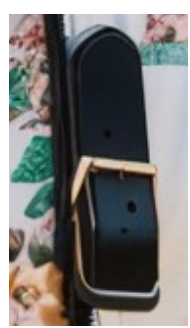

While the massive buckles along the zipper of this jacket may be a design choice, it would be pretty rare.

Structures, objects, and clothes in real life may feature designs that could be considered confusing, atypical, and nonfunctional as well.

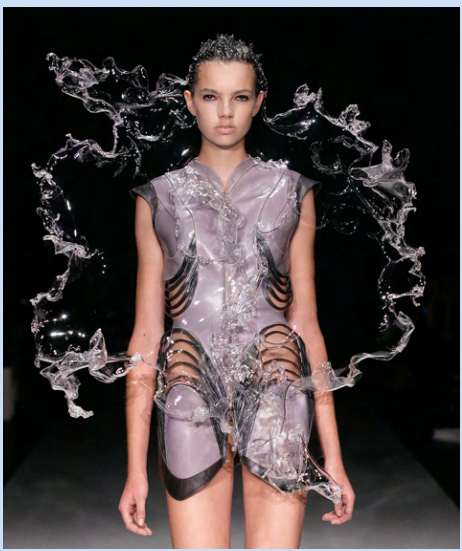

A dress from Iris Van Herpen Spring 2011 Ready-to-Wear collection (Vogue)

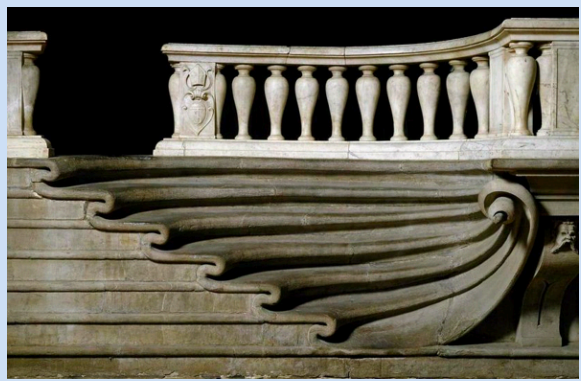

The Marble Staircase of the Presbytery by Bernado Buontalenti for the church of Santa Trinita in Florence, Italy 1574. Photo by Luisa Ricciarini / Bridgeman Images

### 3.4    Text & Logos

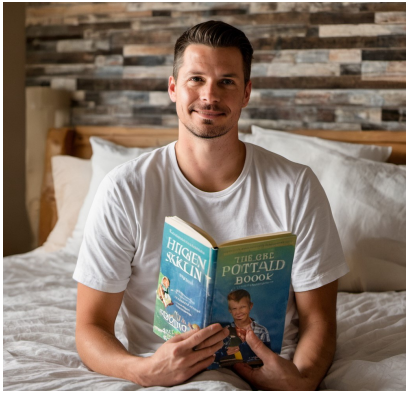

AI-generated text can appear glyph-like, but be in a nonexistent language or produce nonexistent words, spelling errors, and incomprehensible sentences. See if you can read some of the AI-generated text below.

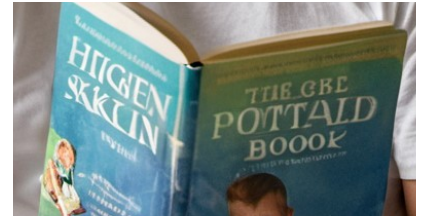

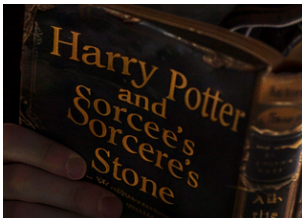

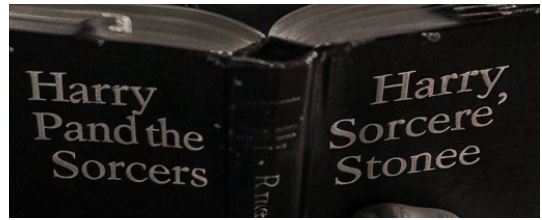

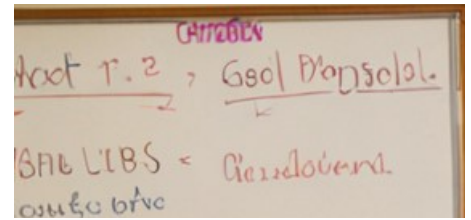

Text in AI image generation is an area that is rapidly improving and these artifacts may not always be noticeable at first glance.

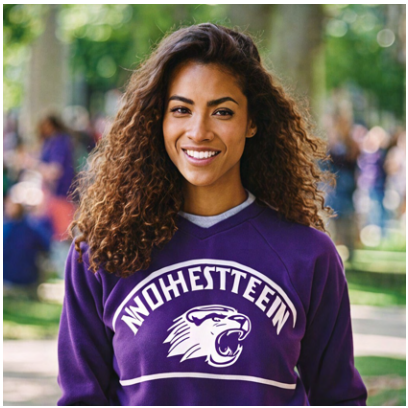

Girl in a Northwestern sweatshirt generated in Stable Diffusion XL (2023). Noticeable distortions and artifacts in the glyphs.

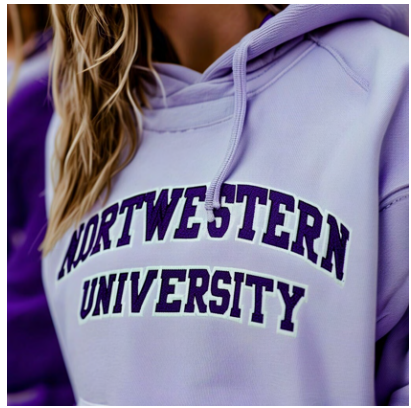

Girl in a Northwestern sweatshirt generated in Stable Diffusion 3 (latest model). Missing "H" in Northwestern.

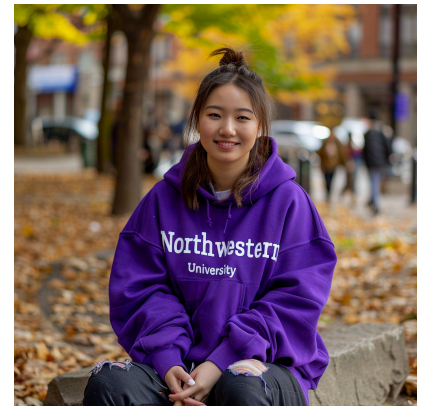

Girl in a Northwestern sweatshirt generated in Midjourney V6 (latest model). Perfect text but an unconventional font for a sweatshirt.

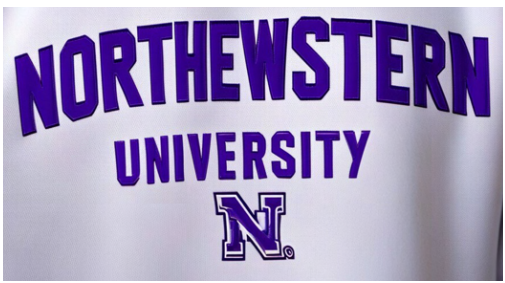

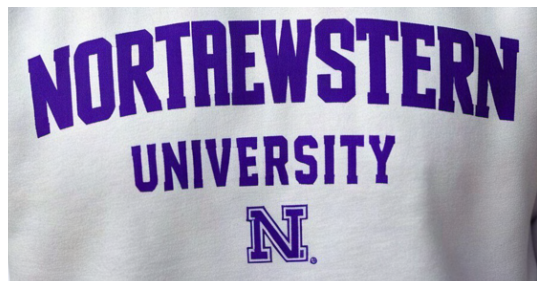

Spelling errors in Stable Diffusion 3.

### 3.5 Prompt Overfitting

AI-generated images may also produce functional implausibilities in the form of prompt overfitting in which certain keywords in the prompt are overrepresented in the image or appear in forms that are uncommon in real life.

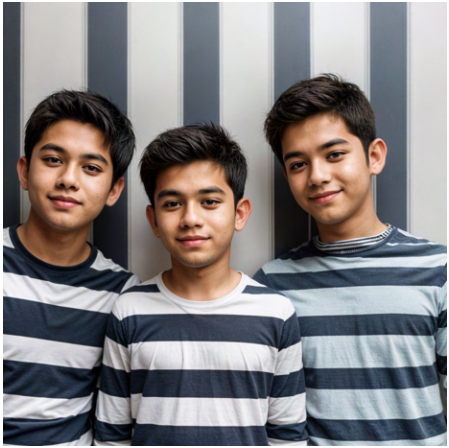

The prompt here was "a group of kids wearing striped shirts." The stripes are overrepresented in the wallpaper behind the group of boys, which was not explicitly stated in the prompt and otherwise would be uncommon.

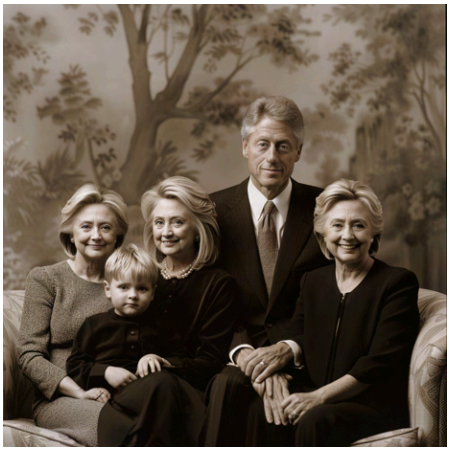
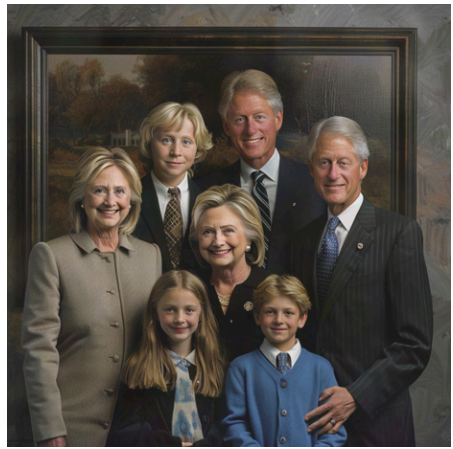

Has Hillary Clinton cloned herself?

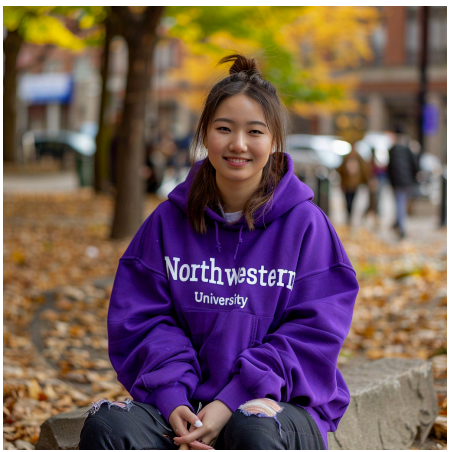
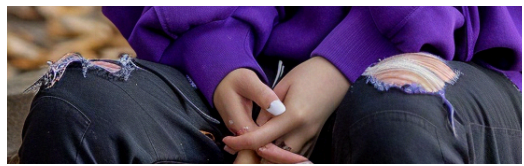

Overrepresentation can be very subtle. If you look closely at the Northwestern student's jeans, the frayed edges are purple, which is an uncommon color to see on black denim. Northwestern colors may be overrepresented here.

## Summary

Functional implausibilities are a distinct artifact in AI-generated images, resulting from a lack of understanding of the fundamental logic of real-world mechanical principles. If the image you are looking at has text, it may be very obvious that it was AI-generated if the text is distorted, has unconventional glyphs, or odd spelling errors. However, functional implausibilities may often be difficult to spot as they are specific to the context of the image. When assessing an image, first look at the objects of the image and see if there are any implausibilities in the objects themselves. Does the oven, watch, broom, or toothbrush look correct? Then, look closely at to how the objects are situated in the environment. Is a person holding it? If so, are they holding it in the right way? Also remember to zoom into the details of objects, particularly clothes and look for any distortion or implausibilities in the small details as well.

Guiding Questions:

- Is the text in the image distorted, does it include unconventional glyphs, or have odd spelling errors?
- Do the objects in the image look right?
- Is an object being held does an object emerge in the setting in an unconventional way?
- Do any of the objects look like they will not function, or are placed in a way that they cannot function?
- Are there any atypical designs, particularly in the prints, buttons, and buckles on pieces of clothing?
- Are there any distortions or glitch-like artifacts in the fine details of objects like the strings of a guitar?

# 04.

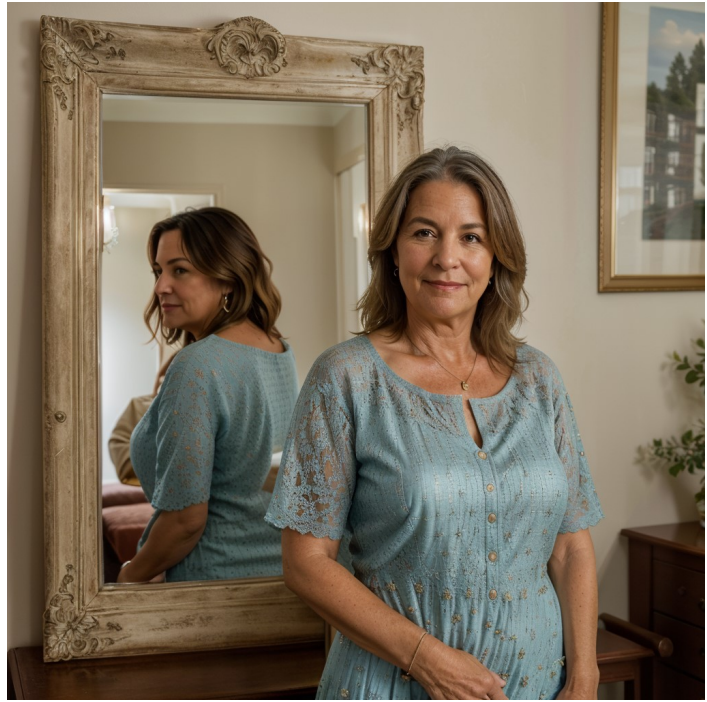

## Violations of Physics

AI-generated images can produce subtle artifacts that are inconsistent with the laws of physics. These include inaccurate shadows, reflections of alternative realities, and depth and perspective issues.

## 4.1 Lighting & Shadows

AI-generated images can produce shadows cast in inconsistent directions or in shapes that do not correspond to their source.

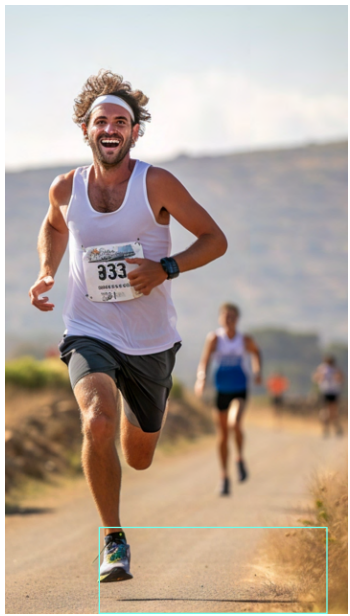
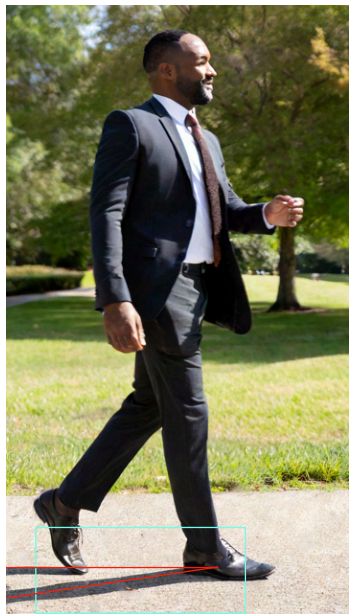
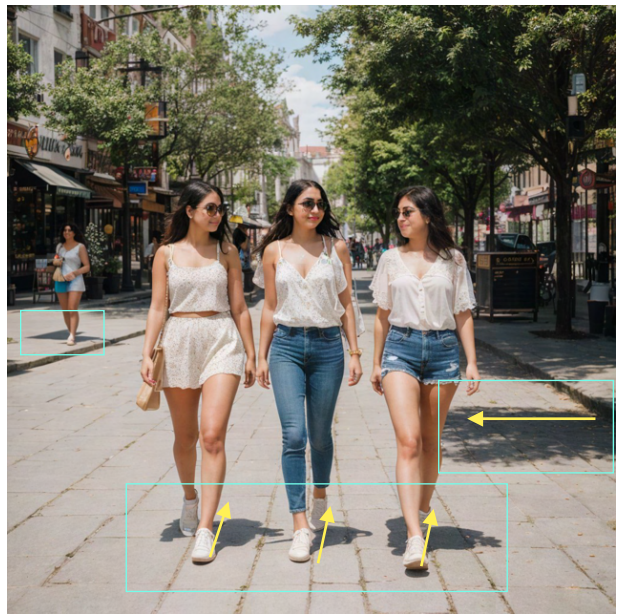

This runner's shadow is a thin strip.

Shadows from a single light source like the sun should be parallel. The leg shadows are not parallel like we would expect it to be.

The shadows of the three girls fall underneath and slightly behind them, but the shadow of the tree on the sidewalk falls to the left of the tree. To be consistent with the shadow of the girls, there should be a tree in the image, in front of the shadow. Additionally, the shadow of the woman in the background mysteriously has sharp corners.

## 4.2 Reflections

Reflections in AI-generated images, whether in mirrors, water, or other specular surfaces may be inconsistent with the rest of the scene.

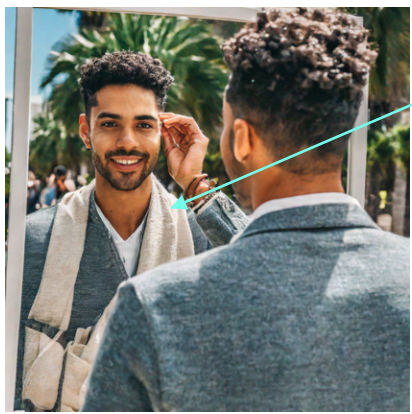

A phantom scarf appears in his reflection.

The man is wearing a short sleeve shirt, but his reflection is wearing a long sleeve shirt that seems to be lighter in color.

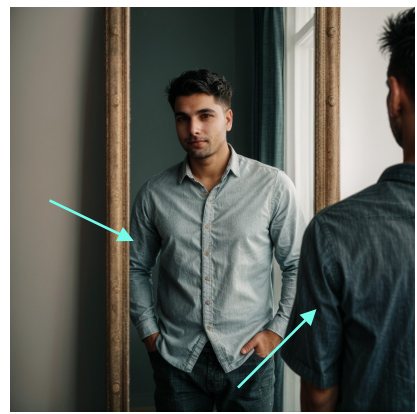

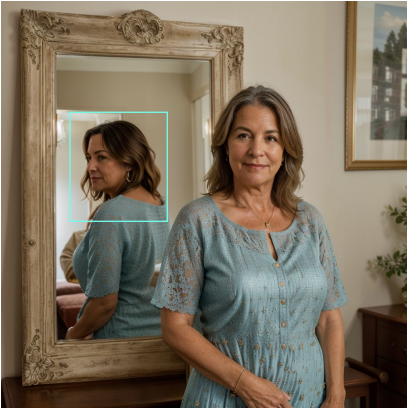
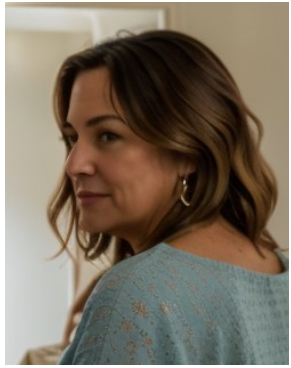

This woman is standing directly in front of the mirror, but her reflection shows her looking backwards.

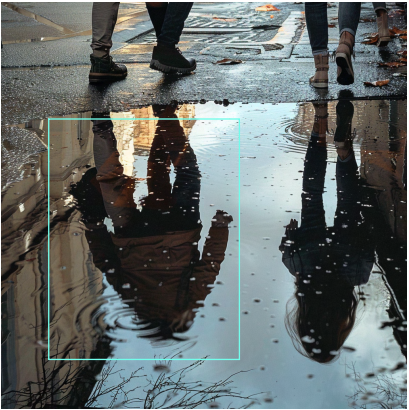
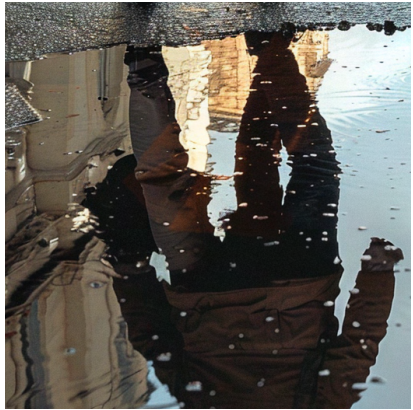

A three legged being appears in the reflection in the puddle.

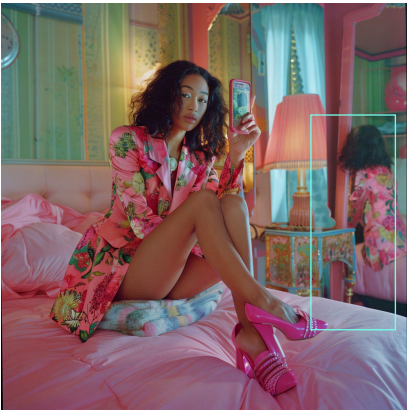
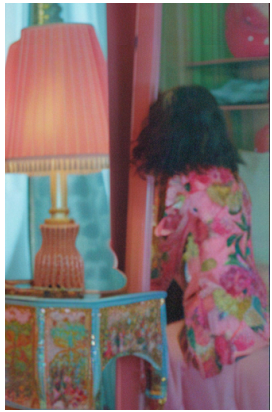

Her reflection in the mirror is facing the opposite direction,

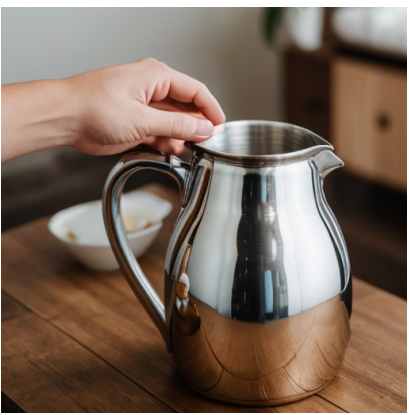
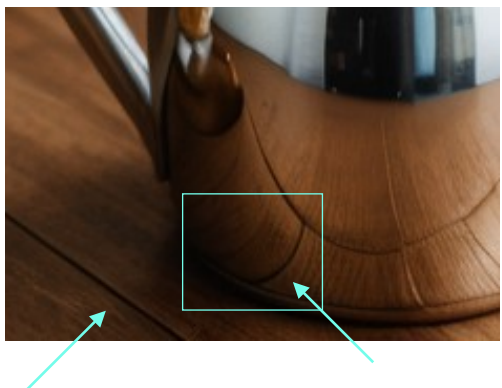

The specular reflection of the table on the metallic jug is not consistent with the lines on the real table.

### 4.3 Depth & Perspective

AI-generated images may produce warping artifacts and depth and perspective issues. These may be subtle and difficult to see, but look carefully for clues in the image that provide some information on the environment of the scene.

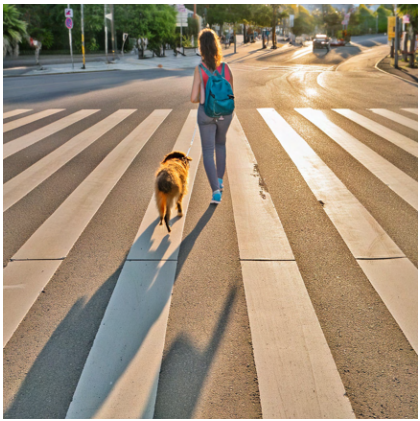

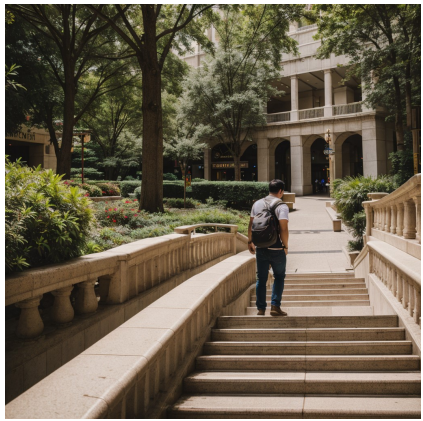

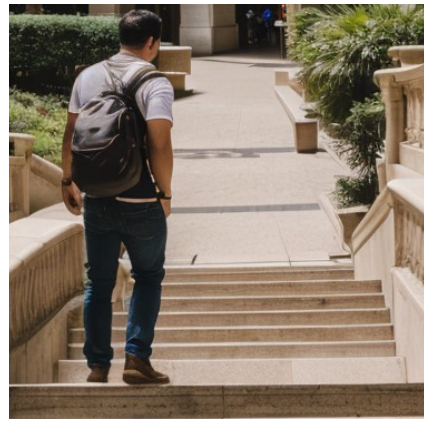

The road converges drastically and then splits off in an unnatural way beyond the end of the crosswalk.

The incline of this staircase does not quite make sense. It appears like the man is on a trajectory that goes both uphill and downhill.

Depth and perspective distortion may also happen in photographs due to different lens focal lengths, which can cause warping of images or change proportions drastically.

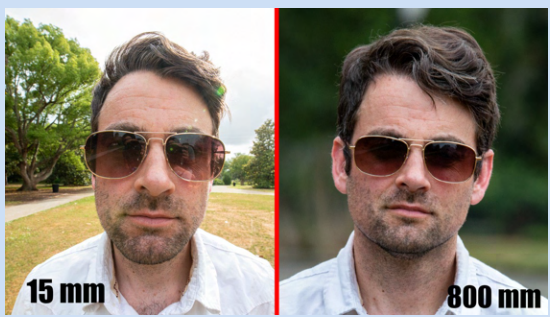

How Lens Compression and Perspective Distortion Work, (FStoppers)

Security cameras often use fisheye lenses that can curve the image.

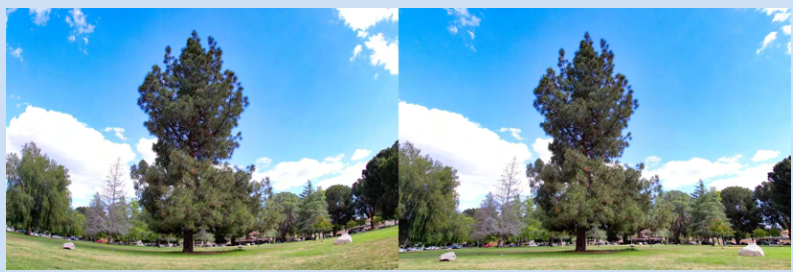

Shot on Fish Eye Lens          Shot on Wide Angle Lens

Fisheye Or Wide-Angle Lens For Travel (Forbes)

## Summary

While we typically take principles of physics for granted, take a moment to look closely at an image and be suspicious of whether the scene is consistent with the obvious physical realities we expect. First, look at any shadows in the image. Be sure to find all of them. Are they consistent with their respective sources? Are they consistent with the shadows in the rest of the image? Depth and perspective issues are less definitive, but make note of anything that seems warped, or a trajectory that does not align with the rest of the image. Are there any straight lines in the image that appear curved? If there are any reflective surfaces in the image like water, mirrors, shiny objects, see if they reflect the world around them. Is this reflected world consistent with the details in the rest of the image? However, keep in mind that camera angles and lens types can distort the image in these ways as well.

Guiding Questions:
- Do any shadows in the image appear inconsistent with their source?
- Do multiple shadows in an image point in different directions?
- Do you notice any warping in the image?
- Does the path or trajectory in a scene appear unnatural or implausible?
- Are there any reflective surfaces in the image? If so, zoom into the reflections - are these reflections consistent with the world around them?

# 05.

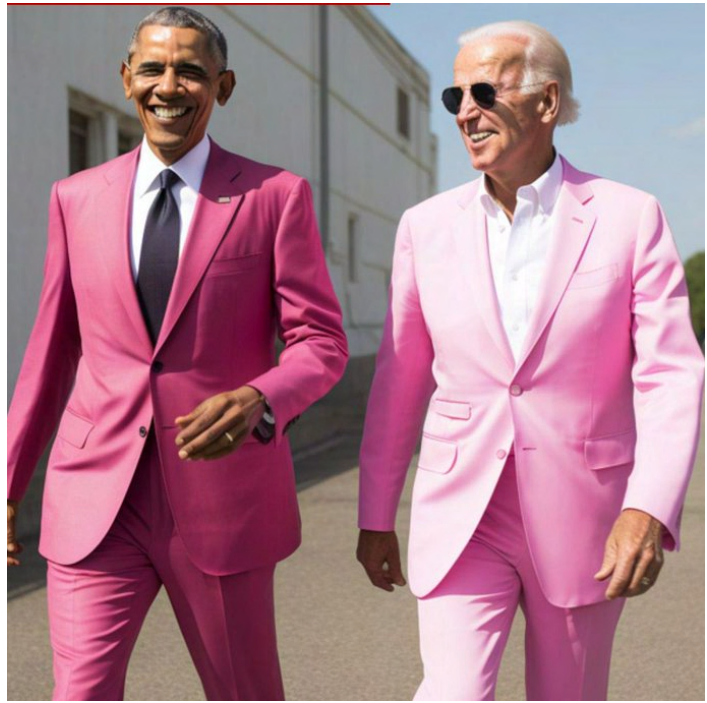

## Sociocultural Implausibilities

AI-generated images may feature scenarios that are inappropriate, unlikely to be seen in the real world, violate subtle rules specific to different cultures, or are historically inaccurate.

**5.1    Unlikely Scenarios**

**5.2    Inappropriate Situations**

**5.3    Cultural Norms**

**5.4    Historical Inaccuracies**

## 5.1 Unlikely Scenarios

AI image generators do not fundamentally understand social context and can produce images that would be unlikely in the real world. These images may be inconsistent with social norms regarding age, environment, cultures, and the behaviors and ideologies of public figures.

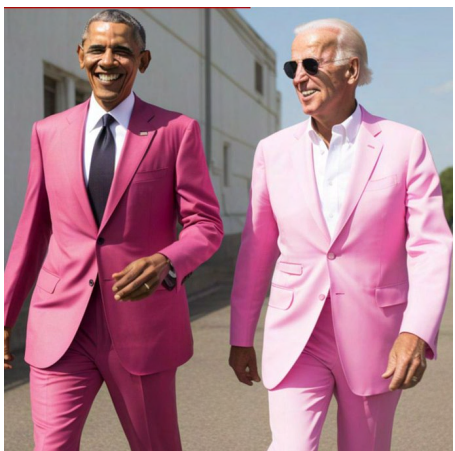

Via Jon Cooper, X

Unlikely for Obama and Biden to be wearing bright pink suits. The image was shared by Democratic fundraiser Jon Cooper.

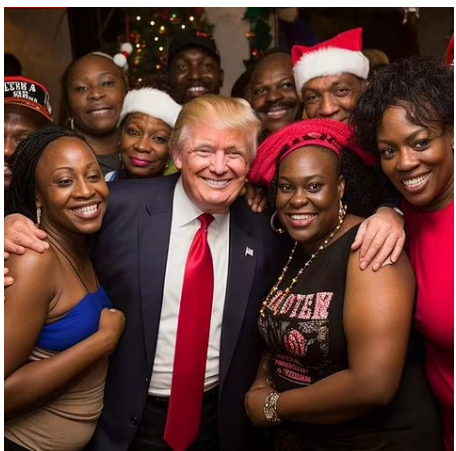

Via Mark Kaye, a conservative radio host in Florida (Newsweek)

This was a deepfake image created by a conservative radio host that went viral. Donald Trump himself shared the image on Truth Social.

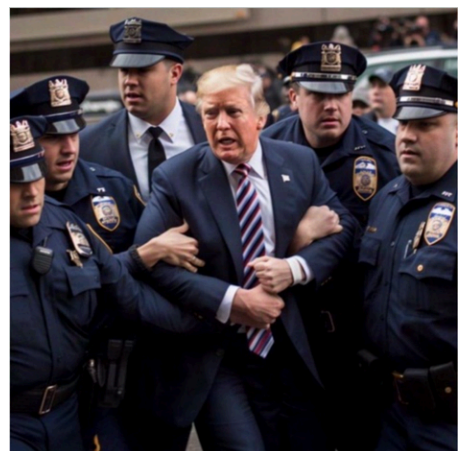

Via Elliot Higgins (Washington Post)

Fake arrest of Donald Trump created by Elliot Higgins, founder of the investigative outlet Bellingcat.

Whether a scenario is unlikely or not is rarely black and white. The following three images are real photos of situations similar to the 3 AI images show above.

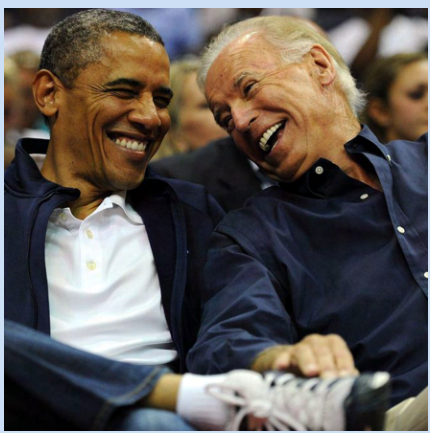

Barack Obama and Joe Biden at a basketball game at the Verizon Center on July 16, 2012 in Washington, DC. Photo by Patrick Smith/Getty Images (Los Angeles Daily News)

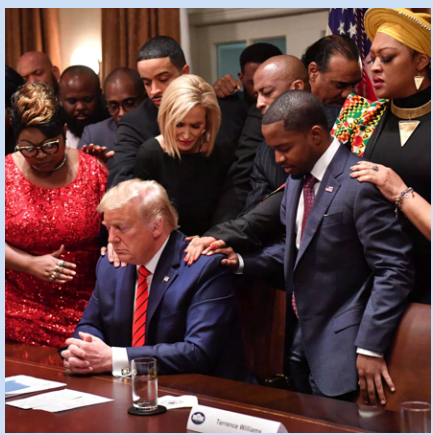

Donald Trump in a prayer circle with African American leaders in the Cabinet Room of the White House on February 27, 2020. Photo by Nicholas Kamm/ AFP via Getty Images (Vox)

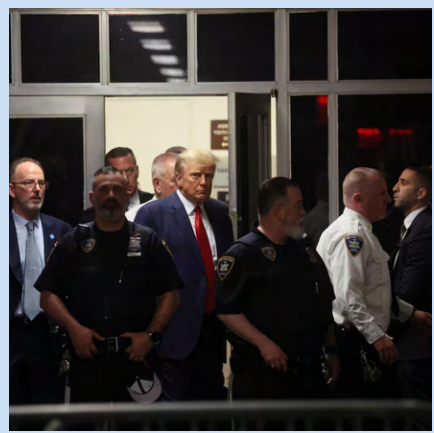

In this real photo of Donald Trump at Manhattan Criminal Courthouse, the police officers do not all have the same face like in the AI-generated fake image. Via *Trump hush money trial* (Reuters)

**05. Sociocultural Implausibilities**

Below are some more examples of unlikely scenarios.

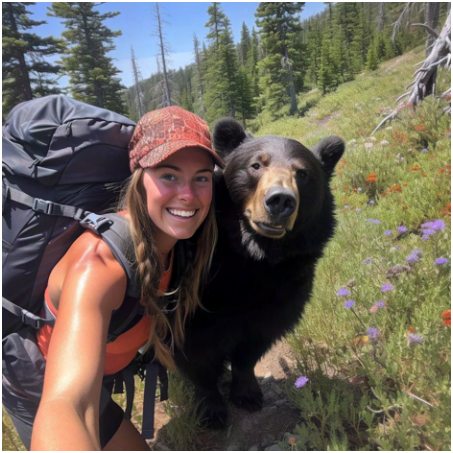

r/midjourney (Reddit)

Selfie with a bear? Quite
unlikely, though possible...

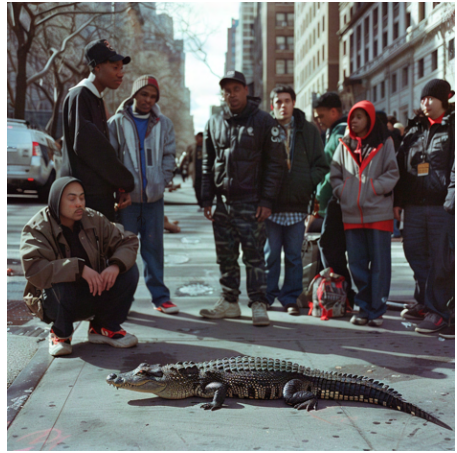

An alligator in New York?

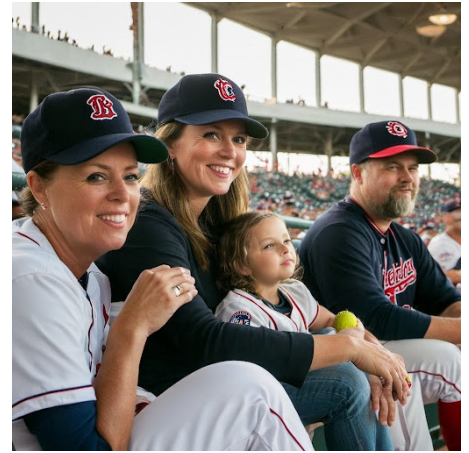

These unlikely scenarios can also
be more subtle like in this image,
where the spectators are wearing
full baseball uniforms including
the pants and socks.

## 5.2   Inappropriate Situations

In addition to unlikely scenarios, AI images may produce scenes that merge details that have diverging
contexts resulting in inappropriate situations.

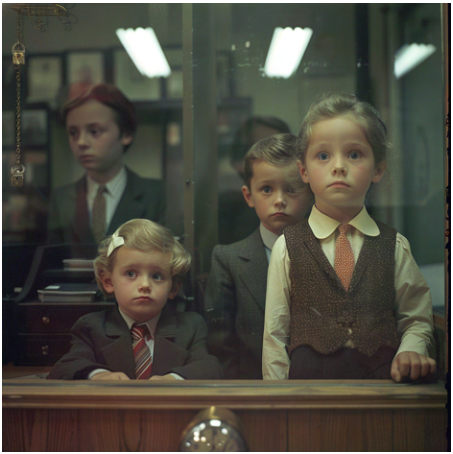

This office appears to be operated
by children dressed from another
era.

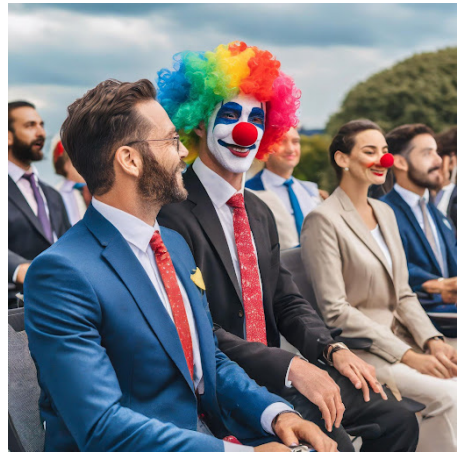

A clown face at a professional
event?

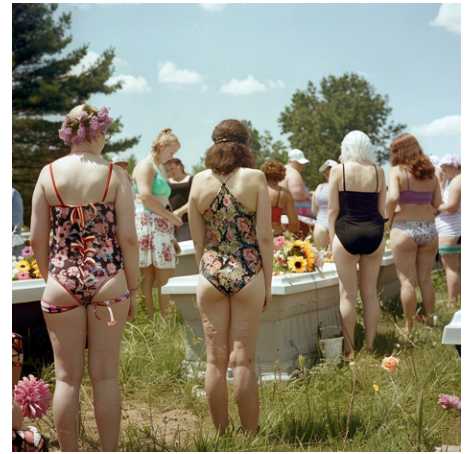

Bathing suits at a funeral?

While a photo with a bear is an unlikely scenario, the following are real images of people with bears.

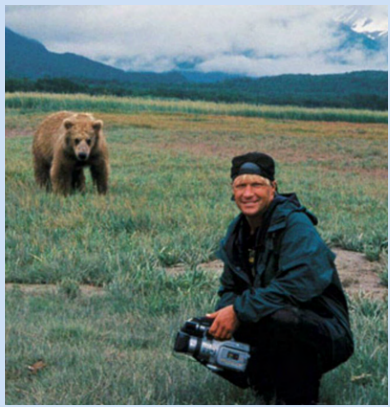

Timothy Treadwell (known as 'Grizzly Man') (Cinematheque)

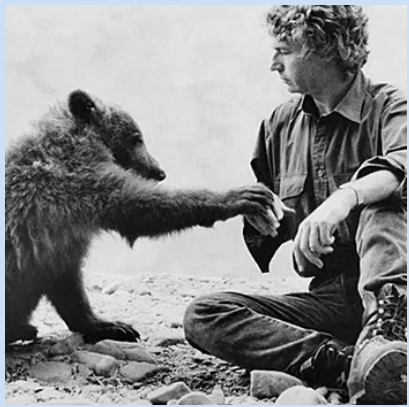

Real trained bears and cubs were part of the 1988 movie *The Bear* (The Bear / L'ours (making of)).

## 5.3    Violations of Cultural Norms

AI image generators may also misrepresent cultural details, as images depicting cultures outside of the West are often fringe cases in the training data. Knowledge of the culture is necessary in order to identify these artifacts, which may reveal themselves in gestures and behaviors that have diverging meanings in different cultures and interactions or clothing that are insensitive or unlikely.

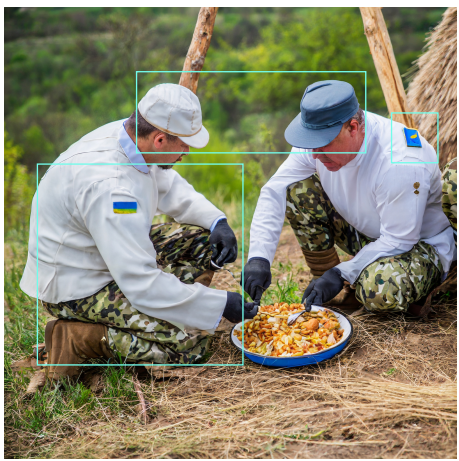

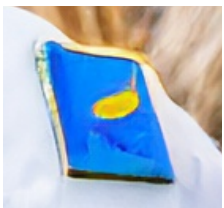

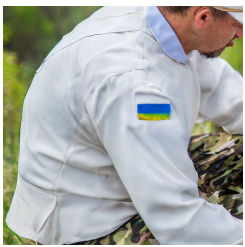

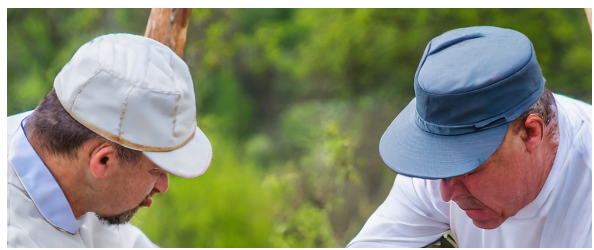

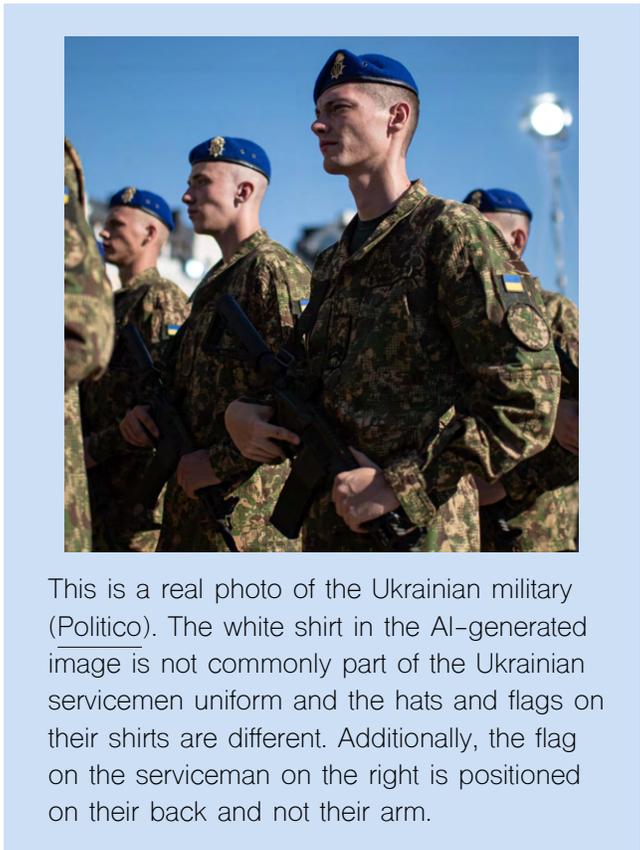

This is a real photo of the Ukrainian military (Politico). The white shirt in the AI-generated image is not commonly part of the Ukrainian servicemen uniform and the hats and flags on their shirts are different. Additionally, the flag on the serviceman on the right is positioned on their back and not their arm.

## 05. Sociocultural Implausibilities

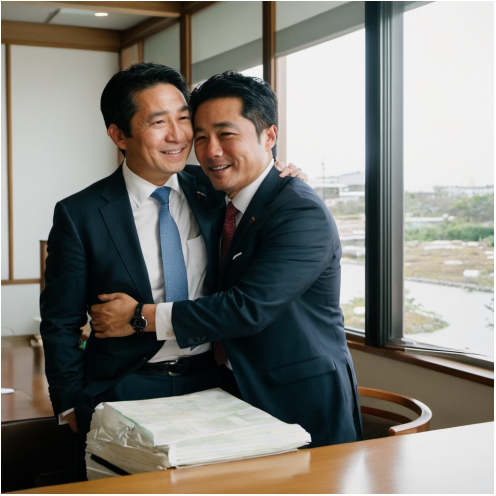
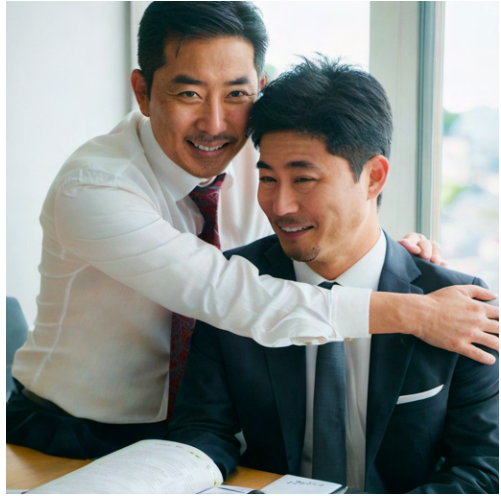

It is uncommon for Japanese people to hug, especially in professional settings.

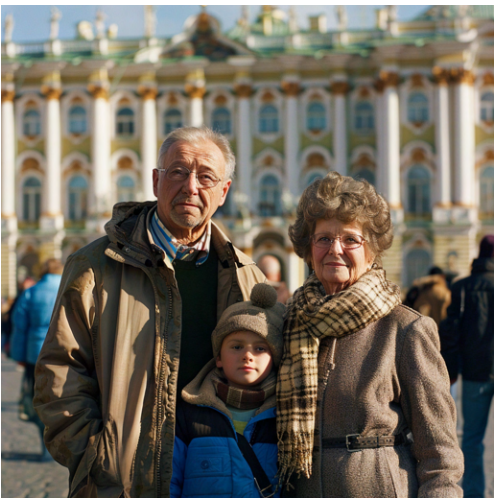
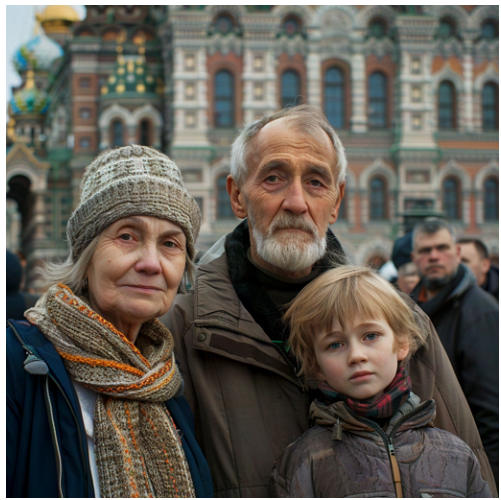

Russians are known to not typically smile towards strangers in photographs.

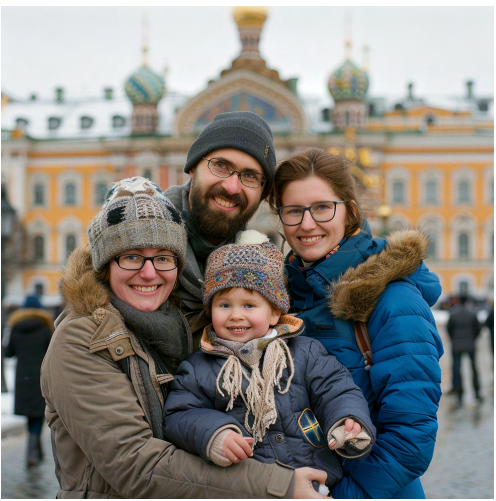
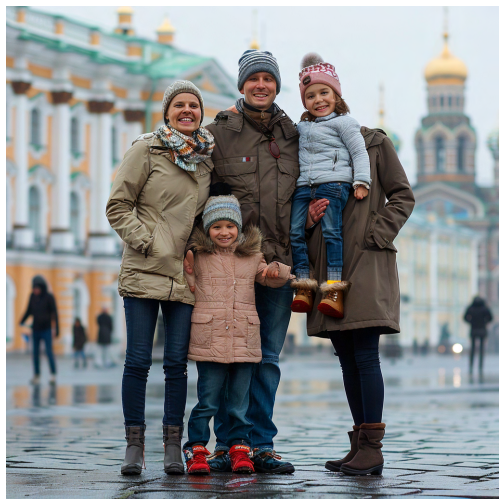

Forcing a smile using an AI image generator results in undergenerated and distorted faces.

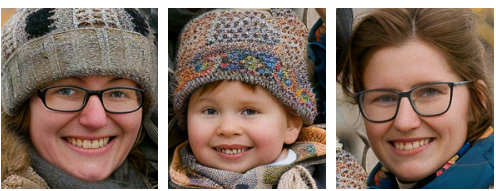
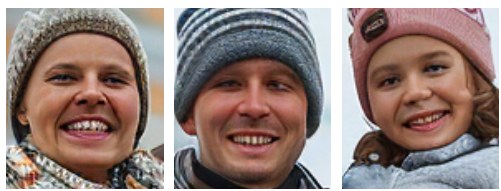

## 5.4    Historical Inaccuracies

AI-generated images may also feature situations from the past that we know are false.

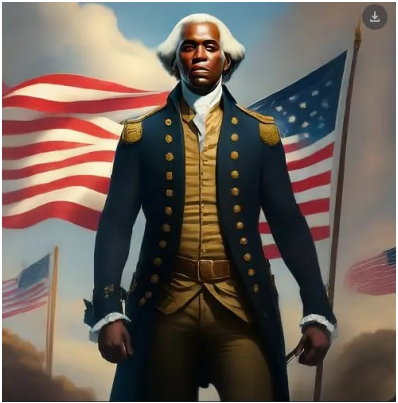

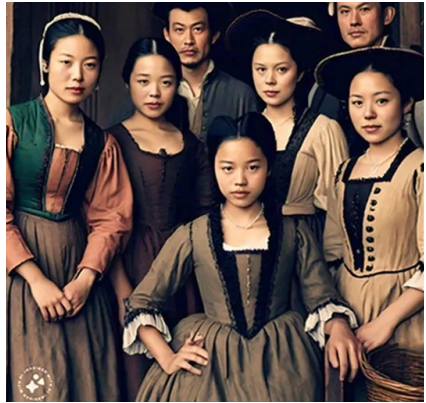

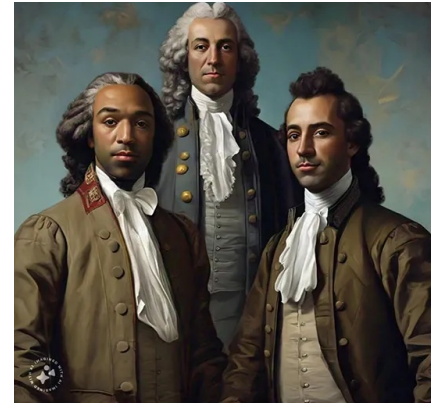

George Washington represented by a black man wearing a white wig from Google's Gemini (NY Post)

A group of people in American colonial times represented by Asian girls by Meta's Imagine image generator (Axios)

Founding fathers represented by people of color by Meta's Imagine image generator (Axios)

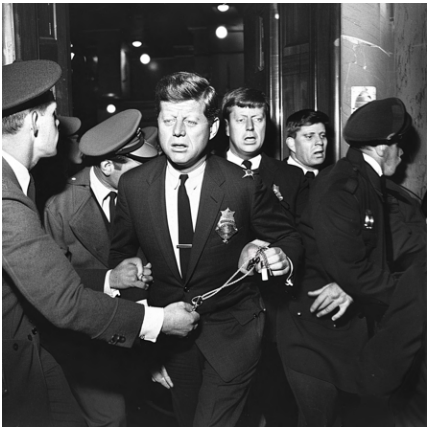

JFK was not arrested.

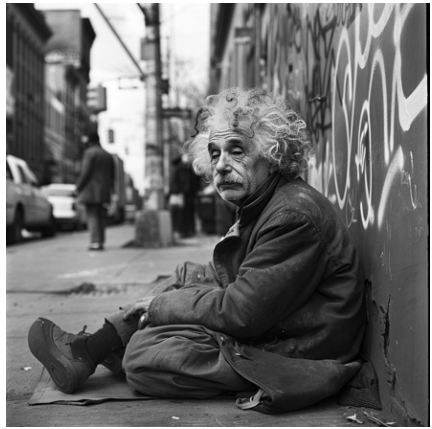

Einstein did not experience homelessness.

## Summary

Sociocultural implausibilities require taking a moment to check in with general common sense and your understanding of cultural context. Is what you're looking at even a plausible scenario? Perhaps it is, but maybe it is very rare like a selfie with a bear or people wearing bathing suits at a funeral. If you notice images of public figures in unconventional settings, check the details on the image or the people with a google search. Cultural implausibilities may be difficult to identify as they potentiality require an in-depth understanding of cultural context. It is impossible for everyone to be an expert in every culture, but if a situation or some details stand out as unconventional, look into it with some research. As what signifies a sociocultural implausibility can be very subjective, it is important to double-check your understanding of the context with other sources.

Guiding Questions:

- Does the image depict an unlikely scenario?
- Does the image show people acting inappropriately or doing things they would not commonly do?
- Does the situation in the image violate the norm in a particular culture?
- If there are public figures in the image, does the image violate a known historical fact?

## Example Walkthrough

Let's try walking through the process of identifying whether an image is AI-generated or not given our understanding of the six categories of visible artifacts in AI-generated images.

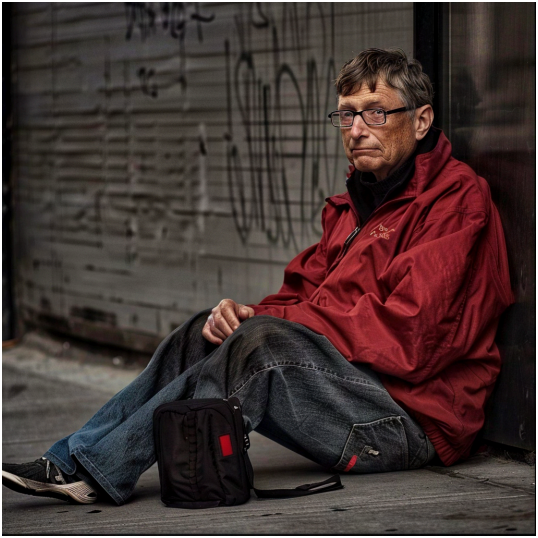

It looks like an image of Bill Gates sitting on the ground looking somber. That would be a pretty rare scene for a photographer to catch so it falls under a **sociocultural implausibility**, but it could be possible so let's continue dissecting the image.

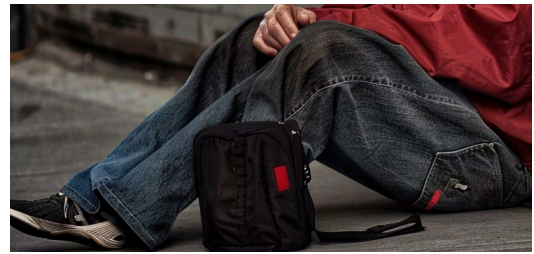

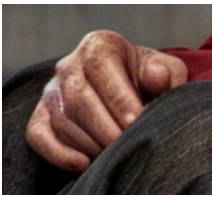

Hands are typically an area where artifacts appear, but there's nothing obviously unnatural with this hand.

Where is his left foot? It looks like his legs merge into one. This is an **anatomical implausibility**.

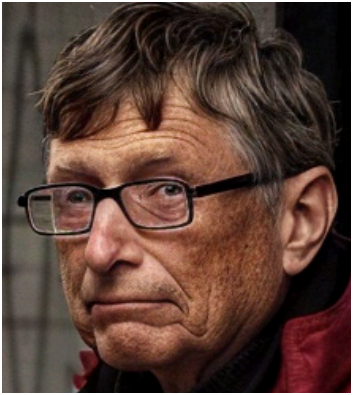

His face overall as well as his hair look slightly glossy and unnaturally radiant in the gloomy scene. This is a **stylistic artifact**.

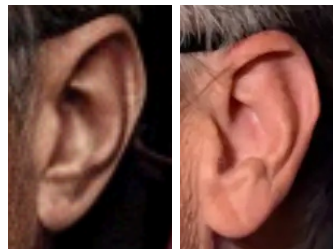

Comparing his ear to a real photo of Bill Gates' left ear reveals some **biometric** differences.

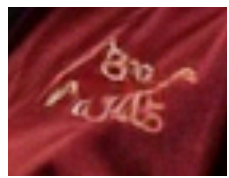

**Atypical design** in the logo/inscription on his chest, but not impossible.

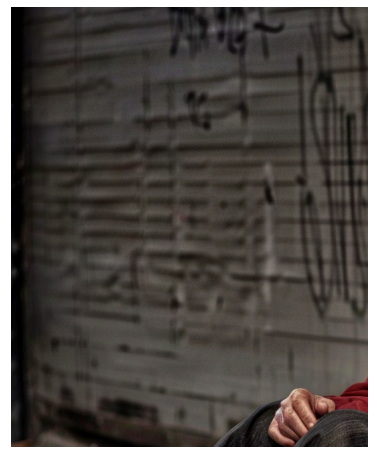

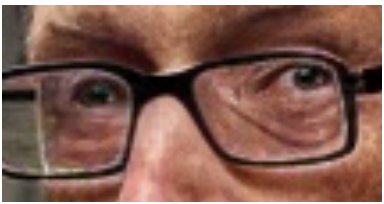

Zooming into his eyes show that his pupils are misaligned - an **anatomical implausibility**. Bill Gates is not known to have a lazy eye.

Garage shutter appears distorted. It is possible it was damaged but this is potentially a **detail rendering** artifact.

# Appendix

## Correlation Between Image Quality and the Prevalence of a Figure/Topic

The more images that are available on a particular topic or figure, the more training data there is for AI image generation models. This results in more common topics and famous public figures to be generated at a higher quality than lesser known topics/people. Let's observe this in images of public figures generated via Midjourney. According to an analysis of Getty Images data in 2024 by Fix the Photo, below are the number of images on Getty Images of some of the most famous public figures:

1. Donald Trump — *cannot be generated
 507,414 images

2. Barack Obama — *intentionally downgraded
353,033 images

3. Queen
Elizabeth II

292,144 images

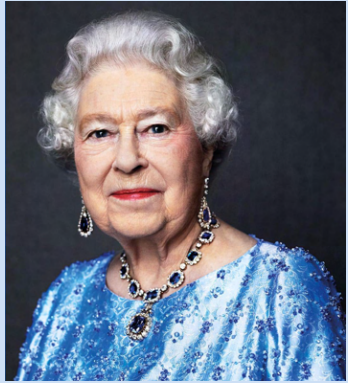
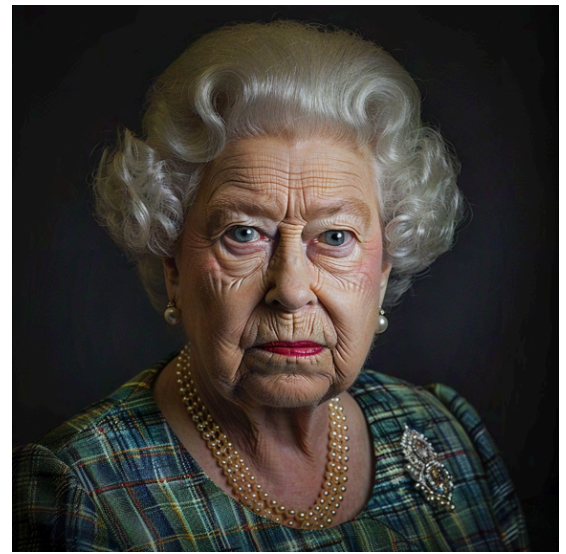

Portrait of Queen Elizabeth by David Bailey (WWD)

6. Kate Middleton

198,227 images

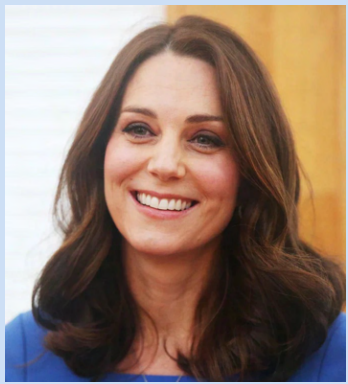
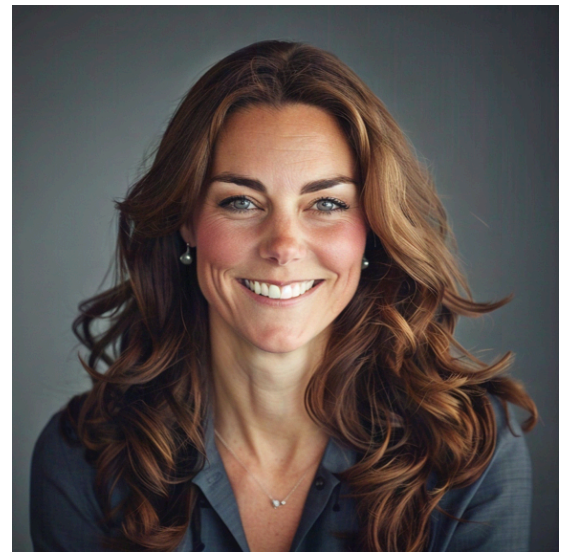

Portrait of Kate Middleton by Jonathan Brady/WPA Pool/Getty (Refinery29)

### 19. Rihanna

96,241 images

Rihanna and Jimmy Fallon at the lower end of the ranking are noticeably worse in quality than Kate Middleton and Queen Elizabeth at the top of the ranking.

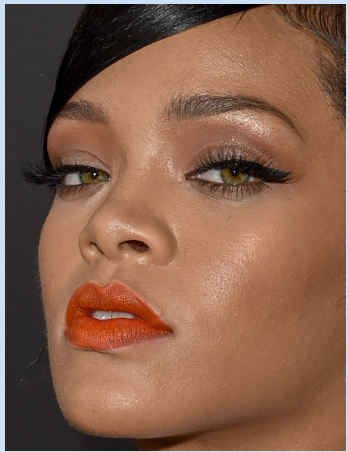
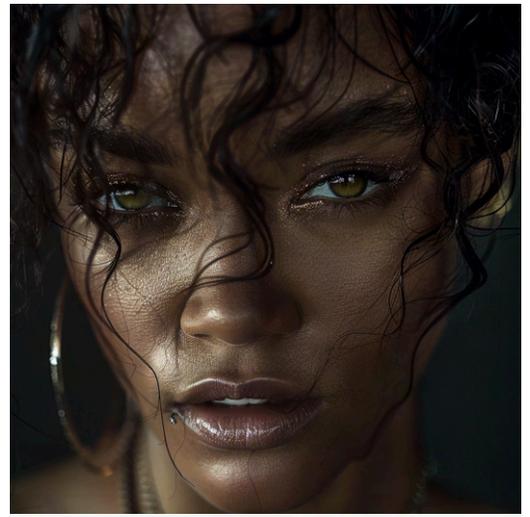

Rihanna face close up by @celebface on X

### 20. Bill Murray

94,083 images

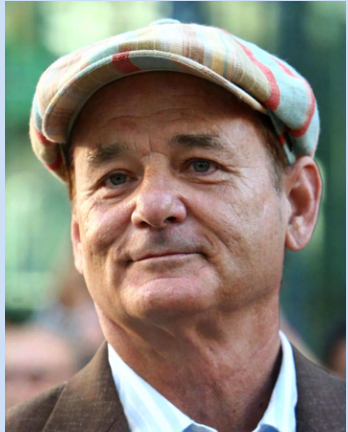
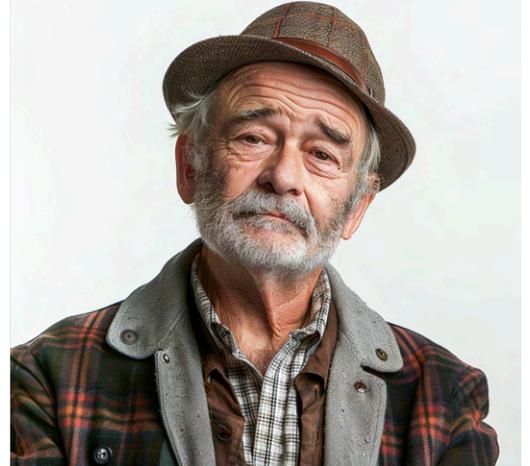

Bill Murray Portrait (Brittanica)

### Nikki Glaser
(unranked)

Comedian Nikki Glaser and actress Golshifteh Farahani are lesser known public figures with a limited online presence. Their generated images portray similar but different people.

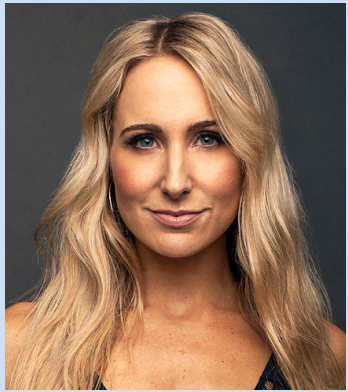
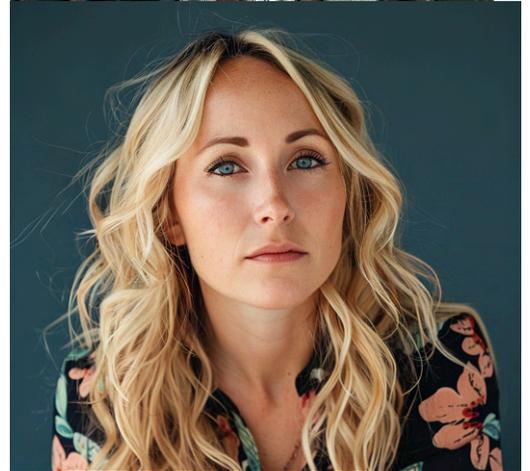

Nikki Glaser Portrait (CityBeat)

### Golshifteh Farahani
(unranked)

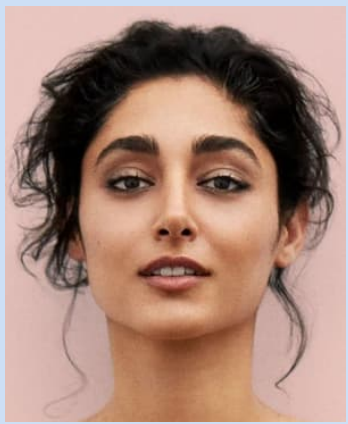
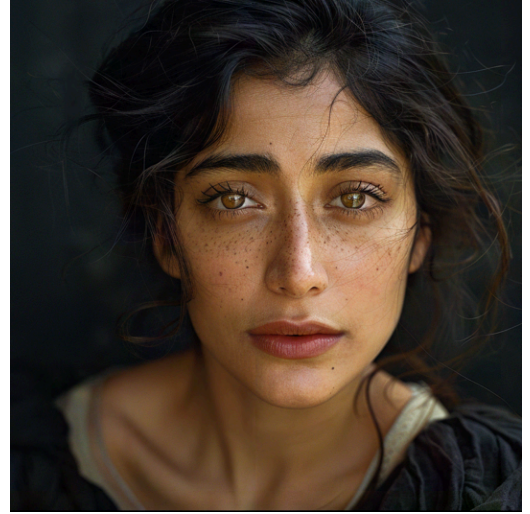

Golshifteh Farahani Portrait (Film Gator)

## Other Generative AI Tools: AI Photo Editing, Image Upscaling & Enhancing

While we have focused on AI-generated images in which the entire image is generated, there are other tools that apply generative AI partially in an image. Images edited through partial applications of AI may be more difficult to spot as there are fewer areas in the image to look out for the artifacts we've discussed. One such tool is generative inpainting, in which only a masked section of an existing image is generated by AI. A specific application of this tool is face-swapped images.

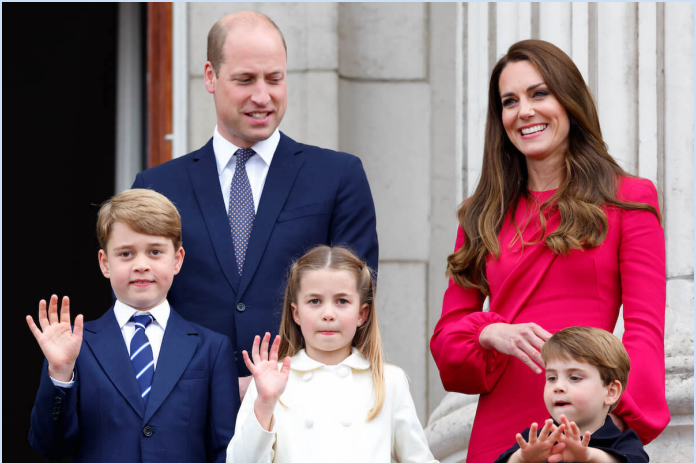

Prince George, Prince William, Princess Charlotte, Prince Louis, and Kate Middleton at the Platinum Pageant in June 2022. Photo by Max Mumby/ Indigo/Getty Images.

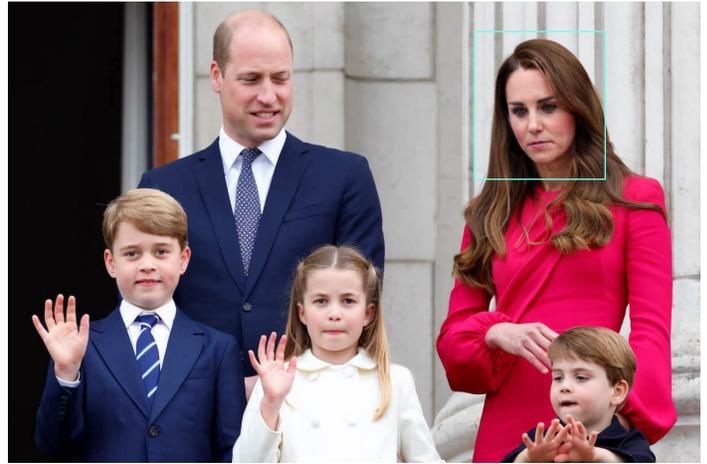

Original image with generative inpainting on Kate's face only.

Generative upscalers and enhancers such as Magnific, Topaz, and Pixelcut take an existing image as the base latent image that the diffusion process generates on top of, allowing the resultant image to maintain most characteristics of the original image, while increasing the resolution. In these upscalers and enhancers, there is often a creativity scale that allows users to select how much an image will diverge from the original input.

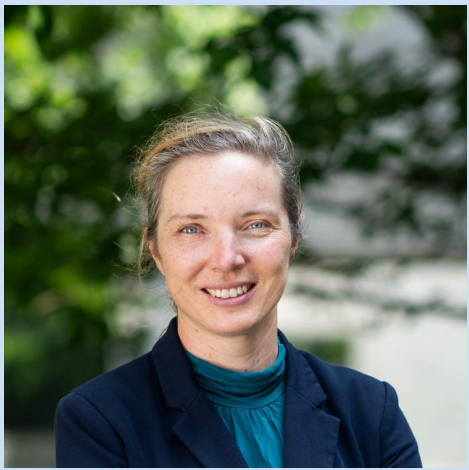

Sigrid Adriaenssens, associate professor of civil and environmental engineering at Princeton University

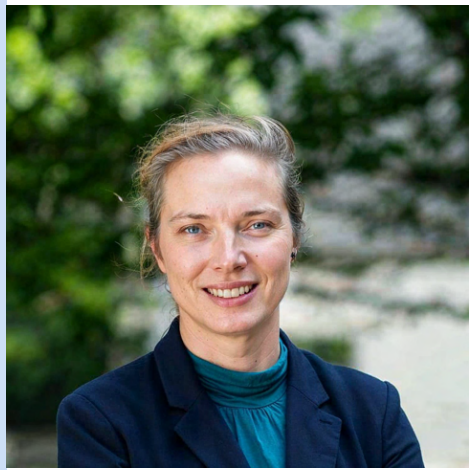

Original image upscaled using Magnific.

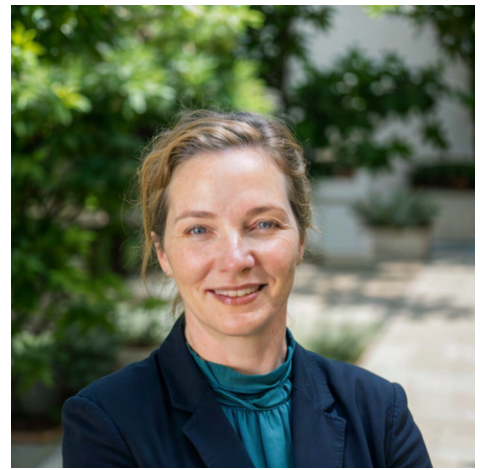

Original image upscaled using a custom Stable Diffusion pipeline.

## Fact-Checking Strategies

With the prevalence of synthetic images on the internet, looking for other sources online for context on images can be an important aspect of determining whether an image is a real photograph or not. The following is a list of some tools and strategies to look for additional information online.

1. Reverse Image Search

Reverse image search allows you to look for sources of a particular image online. This may trace you back to the original source of the image, but there are often multiple occurrences of the same image used in different publications. These publications may have a caption for the image that provides additional detail to the image that has been verified by the publication. Specific tools for reverse image search include:

- Google Images
- TinEye

2. Metadata Analysis

Examining the metadata of an image can provide valuable information about its origin, including the date and time it was taken, the device used, and sometimes even the GPS coordinates. AI-generated images do not have this kind of detailed metadata on how an image was taken. However, keep in mind that metadata can easily be erased and modified through basic digital manipulation such as a screenshot. Metadata is often located in the header of the raw data of the file so a metadata parser is required to read the data. There are many available through an online interface including:

- Online EXIF Viewer

3. Contextual Verification

Developing your own understanding of the the context surrounding an image can help you make a better judgment on whether an image is real of not. Be sure to do some research on the image and the scene portrayed through reputable news sources or official social media accounts. Captions and comments on social media can also provide context. Seek out comments by reporters and researchers in the field who are often on platforms like Twitter to provide brief insight on recent events.

4. Fact-Checking Websites

Several websites specialize in debunking false information and images. These may have already picked up an image you are suspicious of online. Examples include:

- Snopes
- FactCheck.org
- Reuters Fact Check